\documentclass[aps,prx,twocolumn,secnumarabic]{revtex4-2} 

\usepackage[utf8]{inputenc}
\usepackage{longtable}
\usepackage{morefloats}
\usepackage[dvips]{graphicx,color}
\usepackage[dvipsnames]{xcolor}
\usepackage{epsfig,amsfonts,amsbsy}
\usepackage{amsmath,amsthm,amssymb}
\usepackage{bbold}
\usepackage{url}
\usepackage{verbatim}
\usepackage{array}
\usepackage{multirow}
\usepackage{tabularx}
\usepackage{braket} 
\usepackage{dsfont}
\usepackage{bm}
\usepackage{natbib}
\usepackage{multibib}
\usepackage[colorlinks=true, pdfstartview=FitV, linkcolor=blue,
citecolor=blue, urlcolor=blue]{hyperref}
\usepackage[capitalise]{cleveref}

\newcommand{\pGo}{p_{G1}}
\newcommand{\pGt}{p_{G2}}
\newcommand{\pTo}{p_{T1}}
\newcommand{\pTt}{p_{T2}}
\newcommand{\pR}{p_{R}}
\newcommand{\pth}{p^\text{th}}

\begin{document}

\title{Tailoring quantum error correction to spin qubits}
\author{Bence Het\'enyi}
\email{bence.hetenyi@ibm.com}
\author{James R. Wootton}
\affiliation{IBM Quantum, IBM Research Zurich, Switzerland} 
\date{\today}

\begin{abstract}
Spin qubits in semiconductor structures bring the promise of large-scale 2D integration, with the possibility to incorporate the control electronics on the same chip. In order to perform error correction on this platform, the characteristic features of spin qubits need to be accounted for. E.g., qubit readout involves an additional qubit which necessitates careful reconsideration of the qubit layout. The noise affecting spin qubits has further peculiarities such as the strong bias towards dephasing. In this work we consider state-of-the-art error correction codes that require only nearest-neighbour connectivity and are amenable to fast decoding via minimum-weight perfect matching. Compared to the surface code, the XZZX code, the reduced-connectivity surface code, the XYZ$^2$ matching code, and the Floquet code all bring different advantages in terms of error threshold, connectivity, or logical qubit encoding. We present the spin-qubit layout required for each of these error correction codes, accounting for reference qubits required for spin readout. The performance of these codes is studied under circuit-level noise accounting for distinct error rates for gates, readout and qubit decoherence during idling stages.
\end{abstract}

\maketitle

\section{Introduction}

Fault-tolerant quantum computation requires large-scale quantum processors, where quantum states can be encoded in a noise-free subspace of a myriad of noisy qubits. Spin qubits in semiconductor quantum dots offer critical features compatible with having millions of qubits locally connected in a 2D lattice, making them auspicious candidates for fault-tolerant quantum computing~\cite{loss1998quantum,veldhorst2017silicon,burkard2021semiconductor}. Additionally, the half-century-long experience of the semiconductor industry is believed to provide a key advantage over other quantum computing platforms~\cite{veldhorst2017silicon}. Spin qubit platforms to date have demonstrated single- and two-qubit operations and spin readout above $99\%$ fidelity \cite{xue2022quantum,noiri2022fast,oakes2023fast} in devices with up to six qubits~\cite{philips2022universal}. 

Provided that the error rates of the physical qubits are below a certain threshold value, quantum error correction (QEC) codes can suppress errors exponentially with the number of qubits~\cite{knill1998practical,aharonov2008fault,kitaev1997quantum}. One of the most famous QEC codes is the surface code, which requires only nearest-neighbour interaction between qubits on a square grid~\cite{dennis2002topological,bombin2007optimal}. Alongside the modest connectivity requirements, the popularity of the surface code lies on the high threshold error rate and the availability of fast and high-performance classical decoding schemes such as minimum-weight perfect matching (MWPM)~\cite{higgott2022pymatching}. Furthermore, numerous schemes have been developed for the surface code that enable universal quantum computation in a fault-tolerant way~\cite{brown2017poking}. In recent years, several QEC codes have been proposed that fulfill the above-listed criteria \cite{Wootton_2015,srivastava2022xyz,ataides2021xzzx,Hastings2021dynamically,kesselring2022anyon,mcewen2023relaxing}. Some of them even overtake the surface code in terms of error threshold~\cite{ataides2021xzzx} and connectivity requirements~\cite{Hastings2021dynamically,kesselring2022anyon, mcewen2023relaxing}.

The threshold error rate is an important target for qubit platforms, however, its value depends strongly on the noise model assumed to calculate it. In circuit-level error models, individual physical gate errors, qubit readout errors and decoherence during idling are all taken into account~\cite{raussendorf2007topological}. To simplify the model, these errors are assumed to happen with the same probability which makes the resulting numbers hard to compare with physical scenarios where some error mechanisms are more pronounced than others. Moreover, hardware-specific constraints, like noise biases, can significantly change the quantum circuit introducing additional noise channels in the model~\cite{tuckett2020fault}.

Here we provide a detailed quantitative analysis for a wide range of gate and readout fidelities, decoherence and measurement times. Simple formulas are derived to help future experiments estimating their device-specific thresholds and the qubit overhead required for fault tolerance. Our work serves to determine important experimental details required to perform quantum memory experiments. In particular we find fault-tolerant circuits for syndrome measurements and logical state preparation and readout, adapted to each QEC code considered and tailored to the needs of spin-qubit readout.

Moreover, in present readout schemes for spin qubits, defined in single quantum dots, two qubits are required for the readout, one of which acts as a reference spin~\cite{ono2002current}. In such a setting only one qubit of the readout system may be used to encode information. Considering multiple QEC codes, qubit layouts are proposed in this work that acknowledge this constraint. Some codes can  even benefit from this peculiar feature, by reaching a higher effective connectivity, or a sparser grid of qubits that relaxes some requirements for the large-scale fabrication techniques.

This paper is organized as follows in Sec.~\ref{sec:QECspin} the basic concepts of QEC are introduced followed by the derivation of the noise model for spin qubits and some simple formulas are presented relating the parameters of our error model to the experimental figures of merit. In the last subsection we derive the optimal measurement time for QEC applications accounting for qubit decoherence during mid-circuit measurements. These concepts are applied to the surface code in Sec.~\ref{sec:surface} revealing some of the hitherto unexplored parts of the noise-parameter space. Different QEC codes and the corresponding spin-qubit layouts are presented in Sec.~\ref{sec:comparison}, comparing the performance of different codes supplying valuable information about the connectivity qubit-overhead trade-off for future experiments. The main results from which the error-threshold for a device-specific noise model can be deduced are summarized in Tab.~\ref{tab:thresholds}. An outlook for future work and concluding remarks are contained in Secs.~\ref{sec:discussion}-\ref{sec:conclusion}.

\section{Error correction with spin qubits}
\label{sec:QECspin}

\subsection{A brief introduction to quantum error correction}

In this subsection we introduce only the basic concepts and methods in QEC to motivate our choice of the error correction codes and to familiarize the reader with the terminology used in our work. Rather than providing precise mathematical definitions, we instead focus on qualitative explanations for the sake of conciseness. The interested reader may consult a more exhaustive review in the topic such as Refs. \cite{wootton2012quantum,terhal2015quantum}.

Our goal with quantum error correction is to encode logical qubits, i.e., qubits with arbitrarily suppressed error rates, in a low-dimensional subspace of several physical qubits, such that single- and two-qubit gates can be applied between the encoded logical qubits which are read out at the end of the fault-tolerant quantum circuit. One of the prime candidates for such an encoding is the surface code, where it has been shown that all the necessary ingredients can be implemented assuming entangling gates only between nearest-neighbor physical qubits. The rigorous description of some of these methods is beyond the scope of this paper, therefore we resort to quantum memory experiments. In a quantum memory experiment a single logical qubit is encoded in the logical $X$ or $Z$, left idling, while mutually commuting observables are repeatedly being measured, and then read out. Logical errors can be revealed indirectly from the collected observables without having measured the logical qubit.

Since only one qubit of information is needed to be preserved, taking a connected lattice of $N$ physical qubits, $N-1$ independent multi-qubit measurements can be performed, provided that the measurement statistics of the logical subspace remains unaffected. In {\it stabilizer codes}, a set of mutually commuting operators, called {\it stabilizer operators} define these measurements. Logical operators commute with all stabilizers and anti-commute with each other. Stabilizer operators are defined as Pauli strings acting on $\geq 2$ physical qubits. Logical operators act on $\geq d$ qubits, where $d$ is called the {\it code distance}. The code distance defines the smallest number of independent errors that cannot be detected by stabilizer operators, since it is the smallest number of Pauli operators required to perform a logical gate.

Stabilizer measurements ensure that in every cycle the system is projected back into the logical subspace. After several rounds of syndrome measurements, the logical qubit may be read out by measuring every qubit on which the logical operator has a support in the appropriate basis. Since physical errors may or may not have changed the logical information a {\it decoder} is employed, a decoder is a classical algorithm that uses all the stabilizer measurement outcomes (called {\it syndromes}) to guess whether the physical errors have changed the outcome of the logical qubit measurement compared to the encoded information. 

Fault tolerance can be achieved if the decoder has better and better success rate as the number of qubits used for the encoding is increased. The conditions under which this is possible are formulated in the {\it threshold theorem}. The theorem can be formulated in multiple ways, using different assumptions on the noise processes \cite{knill1998practical,aharonov2008fault,kitaev1997quantum}. Let us phrase the statement in the following way: {\it there exists a finite physical error probability $\pth$ of local errors, below which a certain accuracy $\epsilon$ of the quantum computation of depth $D$ can be ensured by encoding logical qubits in $\text{polylog}(D/\epsilon)$ physical qubits.} Summarizing the arguments above, the threshold error rate $\pth$ of the physical qubits depends on three factors, {\it (i)} the error correction code itself, {\it (ii)} the decoding algorithm, {\it (iii)} and the error model. 

Here we focus on a special class of stabilizer codes which is defined by the topological properties of the error syndromes which is easiest to understand in the Hamiltonian formalism. A Hamiltonian can be defined from the negative sum of all stabilizer operators. The ground state of this Hamiltonian (i.e., the mutual $+1$ eigenstate of all stabilizers) is two-fold degenerate. Logical operators act on this subspace as gauge operators. Single qubit errors change some of the syndromes, creating point-like excitations above the ground state. If the possible excitations in the model are composed of two bosonic paticles which are their own anti-particle and have non-trivial braiding statistics, the model is called a $D(\mathbb Z_2)$ anyon model~\cite{kitaev2003double}. Throughout this work we will refer to this set of codes as the {\it surface code family} after its most prominent representative, the surface code.

The choice of the decoder has a large impact on the error threshold. The maximum-likelihood decoder is shown to result in an optimal decoding, but has an exponential computational complexity which makes it unfeasible for system sizes large enough for fault tolerant quantum computations especially if the syndrome information needs to be processed in real time~\cite{battistel2023realtime}. For members of the surface code family, error syndromes always appear pairwise due to the general properties of the underlying anyon model. This crucial property makes these codes amenable for decoding by minimum-weight perfect matching (MWPM) which is almost linear computational complexity \cite{higgott2022pymatching}. Together with the qualitative similarities to the well-established surface code, this motivated our choice to focus on the surface code family and use the same MWPM decoding scheme throughout our work.

The error model incorporates probability of different type of errors, as well as spatial and temporal correlations between error events. In the next subsections we will consider the physical origin of noise in spin qubit devices and derive an error model that acknowledges the special properties of this platform. The error model needs to be efficiently simulable, such that one can make predictions about the resources required for fault-tolerant quantum computing.

\subsection{Noise model of spin qubits}
\label{sec:noisemodel}

\begin{figure*}
    \includegraphics[width = \textwidth]{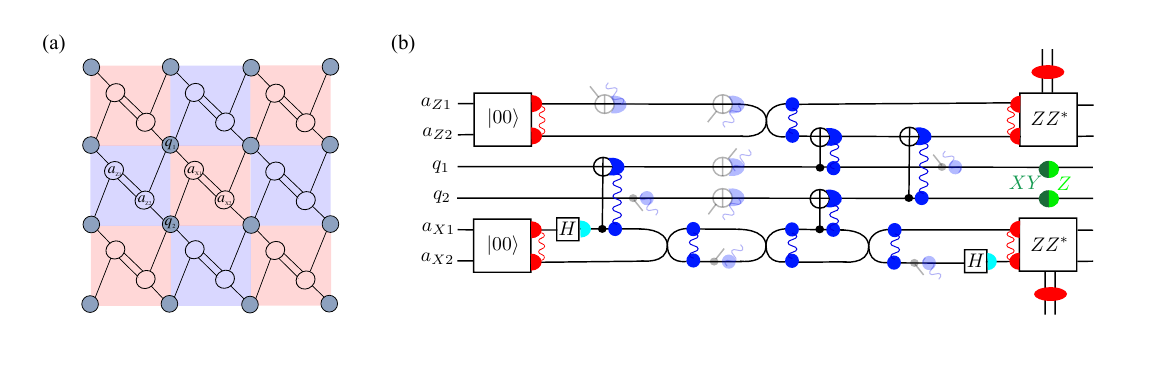}
    \caption{(a) Qubit layout and connectivity of a bulk section of the surface code. Gray filled nodes represent qubits that are connected to ancillas (empty nodes) via single links representing connections via two-qubit gates. Ancillas can be read out pair-wise or swapped with each other (double link). $X$ and $Z$ stabilizer plaquettes are shown in red and blue respectively. (b) Circuit representation of qubits $q_1$ and $q_2$ taking part in an $X$ and a $Z$ stabilizer measurement. The qubits shown are also part of two additional stabilizer measurements each (implied by faint gates). The CX schedule and the connectivity necessitate three (one) SWAP gates for the $X$ ($Z$) stabilizer measurements. Red, cyan, and blue colored elements in the circuit represent single- and two-qubit depolarizing noise (or bit flip on classical lines) for readout, single-, and two-qubit gates, respectively. During the measurement qubits experience $X$-$Y$ (dark green) and $Z$ errors (light green) with different rate.}
    \label{fig:SurfCodeLayoutCircuit}
\end{figure*}

In this section we consider some common features and constraints of spin-qubit devices and use them to establish the error model to be used for the calculation of the threshold surface of the surface code and its comparison to other QEC codes of the same family. There are multiple ways to encode qubits in the spin and orbital degrees of freedom of semiconductor quantum dots \cite{burkard2021semiconductor}. Here we focus on the type of spin qubits, where the qubit states correspond to the spin projections of a single electron or hole in a quantum dot that is split by a magnetic field according to the original proposal of Loss and DiVincenzo~\cite{loss1998quantum}. 

Arbitrary single qubit rotations as well as the two-qubit CX gate can be implemented natively in with spin qubits~\cite{philips2022universal}. Gate errors are often characterized with a single number the gate fidelity $\mathcal F_G$. As opposed to gate tomography, the fidelity does not give detailed information about the noise channels. Since different spin-qubit platforms can have very different noise channels, we will focus on the gate fidelity and assume that single- and two-qubit gates are followed by single- and two-qubit depolarizing noise, i.e., the $m$-qubit density matrix becomes $\rho_{m} \rightarrow (1-p_{Gm}) \rho_{m} + 2^{-m}p_{Gm} \mathbb{1}_{2^m}$, where the probabilities are determined by the fidelity as
\begin{subequations}
\begin{equation}
    \pGo = 1-\mathcal F_{G1} \equiv p_G \frac{2}{1+\eta_G},
\end{equation}
\begin{equation}
    \pGt = 1-\mathcal F_{G2} \equiv p_G \frac{2\eta_G}{1+\eta_G},
\end{equation}
\end{subequations}
with $p_G$ being the average gate fidelity and the error bias between single- and two-qubit gates reads
\begin{equation}
    \eta_G = \frac{\pGt}{\pGo}.
\end{equation}

An idling spin qubit loses its phase coherence at a higher rate than the change in population of its basis states. Dephasing can come from the low-frequency fluctuations of the qubit splitting that changes the precession frequency of the spin around the axis of the magnetic field. Such low-frequency noise often originates from the slow dynamics of nuclear spins, or $1/f$ charge noise coupling to the spin splitting via spin-orbit interaction~\cite{coish2004hyperfine,culcer2009dephasing}. Relaxation, on the other hand is dominated by phonon emission which is typically suppressed by the low phonon density of states at the energy of the qubit splitting~\cite{tahan2014relaxation}.

Relaxation can be treated in the Bloch-Redfield approximation \cite{bloch1957generalized}, which yields an exponential decay of the diagonal elements of the density matrix. This is equivalent to Pauli X and Y errors (i.e., $\rho \rightarrow (1-p_{T1}) \rho + \frac{p_{T1}}2 (X\rho X + Y\rho Y)$) with probability
\cite{tomita2014lowdistance}
\begin{equation}
    p_{T_1} = \frac{1-e^{-\tau_i/T_1}}4 \approx \frac{\tau_i}{4T_1}.
\end{equation}
with $T_1$ being the relaxation time and $\tau_i$ is the idling time. On the other hand, dephasing due to low-frequency noise is better described in the filter function formalism \cite{burkard2021semiconductor}, leading to a Gaussian decay of the off-diagonal element of the density matrix with a time scale $T_2$. This process corresponds to Pauli-Z errors (i.e., $\rho \rightarrow (1-p_{T2}) \rho + p_{T2} Z\rho Z$) with probability 
\begin{equation}
    p_{T_2} = \frac{1-e^{-\tau_i^2/T_2^2}}2 - \frac{1-e^{-\tau_i/T_1}}4 \approx \frac{\tau_i^2}{2T_2^2}-\frac {\tau_i} {4T_1}.
\end{equation}
The difference in decoherence rates can be quantified with the noise bias 
\begin{equation}
    \eta_T = \frac{p_{T_2}}{p_{T_1}} \approx \frac{2\tau_i T_1}{T_2^2} - 1,
    \label{eq:etaT}
\end{equation}
while the total idling error rate reads as $p_T = p_{T_1} + p_{T_2}$. However, we note that $\eta_T \ll T_2/T_1$ (since $\tau_i \ll T_1,T_2$), meaning that the noise bias can be substantially lower than the naive expectation of $\eta_T \sim T_2/T_1$. Furthermore, one might apply an arbitrary dynamical decoupling pulse sequence during idling further improving the $T_2$ dephasing time that enters the above equations. Using these considerations we expect $\eta_T \sim \mathcal O (10)$~\cite{noiri2022fast}.

\begin{table*}[htp]
\begin{center}
\begin{tabular}{|c|c|c|c|c|c|c|c|}
\hline
\rule{0pt}{10pt}  & connectivity & $\nu_q/\nu_a$ & $p_G^\text{th}[\%] $ & $p_T^\text{th}[\%]$ & $\pR^\text{th}[\%]$ & $\sqrt{\braket{\left(\frac{\delta \pth}{\pth}\right)^2}}$ \\[2pt]
\hline
\rule{0pt}{10pt} Surface code & $3\tfrac 1 3$ & 1/2 & $0.82$ & $3.94$ & $14.5$ & 0.08\\[2pt]
\hline
\rule{0pt}{10pt} XZZX code & $3\tfrac 1 3$ & 1/2 & $0.37$ & $15.1$  & $12.9$ & 0.11\\[2pt]
\hline
\rule{0pt}{10pt} 3-CX surface code & $2\tfrac 2 3$ & 1/2 & $0.65$ & $4.1$ & $8.7$ & 0.1\\[2pt]
\hline
\rule{0pt}{10pt} XYZ$^2$ code & 4 & 2/2 & $0.465$ & $4.2$ & $16.5$ & 0.12\\[2pt]
\hline
\rule{0pt}{10pt} Floquet color code* & $2\tfrac 1 4$ & 1.5/4.5 & $0.48$ &$ 0.7$ & $1.41$ & 0.005\\[2pt]
\hline
\rule{0pt}{10pt} Honeycomb Floquet code* & $2\tfrac 1 4$ & 1.5/4.5 & $0.43$ & $1.16$ & $0.99$ & 0.015\\[2pt]
\hline
\end{tabular}
\end{center}
\caption{Spin-qubit connectivity (i.e., average number of two-qubit connections in a 2D lattice) and the parameters of the linearized threshold surface appearing in Eq.~\eqref{eq:poverpth}. The total number of qubits and ancillas are given by $N_{q,a} = \nu_{q,a} d^2$ for a code distance $d$. The overall performance of the fitting is further characterized by the maximum and the mean deviation from the numerical value, i.e., $\delta \pth = \pth - p^\text{num,th}$. Thresholds are given at $\eta_T = 20$ and $\eta_G =1$. Asterisk denotes syndrome measurement via Bell-state preparation.}
\label{tab:thresholds}
\end{table*}

Readout of spin qubits is typically carried out via spin-to-charge conversion and charge sensing. Here we only consider those conversion schemes which do not require a reservoir connected to the quantum dot accommodating the qubit. The conversion of spin to charge involves two spin qubits in close proximity, one in a known state and another one to be measured. Reducing the tunnel-barrier between the two quantum dots gives rise to a spin-selective tunneling, i.e., Pauli-spin blockade, after which the (change in) charge state of the quantum dot can be detected. Alternatively, one can exploit the strong spin-photon coupling in a setup where a single particle with strong spin-orbit interaction is situated in a double quantum dot that is coupled to a resonator. From an architecture point of view both of these approaches require twice the space of a single qubit. Therefore, in the following we assume that readout involves pair of qubits in the qubit layouts to be presented.

The equipment for charge sensing imposes further restrictions on the qubit architecture. Charge sensing can be carried out using single-electron transistors (SET) in a radio-frequency reflectometry setup, which is a popular method for small-scale devices. However, this requires an SET in the proximity of the spin qubits which has a substantial footprint compared to the quantum dot dimensions. Scalable architectures can potentially employ gate-dispersive sensing techniques with signal multiplexing, similar to that of superconducting qubits, relaxing the spatial requirements of the sensor near the qubits~\cite{vigneau2022probing}.

An important aspect of the readout is the compromise to be made between measurement time $\tau_R$ and fidelity. The fidelity of charge sensing is exponentially improved by increasing the measurement time~\cite{vigneau2022probing}. That does not mean that the longer one measures, the better it gets. Measurement time is limited by relaxation processes (deep in the Pauli blockade for charge-sensing) or Landau-Zener tunneling (in the case of a dispersive readout). The readout error can be described (qualitatively) as
\begin{equation}
    p_R(\tau_R) = 1 - \mathcal F_R(\tau_R) = 1 - (1-e^{-\tau_R/\tau_\text{min}})e^{-\tau_R/T_{1R}}
    \label{eq:ROerrorrate}
\end{equation}
where $T_{1R}$ is the decoherence time of the qubit being read out and $\tau_\text{min}$ is the minimum integration time, i.e., time needed to achieve a signal-to-noise ratio of 2~\cite{vigneau2022probing}.
The maximum readout fidelity is then achieved at the measurement time
\begin{equation}
    \tau_R^* = \tau_\text{min} \log \left(1+\frac{T_{1R}}{\tau_\text{min}} \right).
    \label{eq:tRexp}
\end{equation}
In our error model readout errors are two-qubit depolarizing errors followed by a classical bit flip on the measurement outcome (e.g., infidelity of the charge detector). Both of these lead to a faulty syndrome bit with a joint probability $p_R$. 

Finally, we note that within the Pauli noise model, reset errors during the stabilizer measurements generate the same syndrome as readout errors. Therefore, we merge these error rates into a single error parameter, the reset-readout error rate
\begin{equation}
    p_{RR} = p_\text{res}(1-p_R(\tau_R)) + p_R(\tau_R) (1-p_\text{res}),
\end{equation}
where $p_\text{res}$ is the probability that a faulty initialization flips the measurement outcome (e.g., for depolarizing noise $p_\text{res} = \frac 8 {15} (1 - \mathcal F_\text{init})$).

\subsection{Error-threshold surface and resource estimation}

If the ratio of different probabilities is kept constant, one obtains a single-parameter error model where $p=0$ corresponds to no errors. Therefore, according to the threshold theorem a finite error threshold $\pth$ can be found, the value of which will depend on the ratio of error probabilities. Asymptotically, for $p\ll \pth$ the logical failure rate can be approximated as
\begin{equation}
    P_L \sim w \left( \frac{p}{\pth} \right)^{d/2},
\end{equation}
where $w$ is a prefactor that depends on how many length-$d/2$ paths can lead to logical failure. Considering different ratios of the error parameters, i.e., different directions in the error-parameter space, and calculating the corresponding threshold values maps out a threshold surface in the parameter space $\mathbf p$ that encloses the origin ($\mathbf p = 0$). Inside the threshold surface the logical error rate can be decreased arbitrarily by increasing the number of qubits. 

Fixing the two error bias parameters $\eta_G$ and $\eta_T$, we are left with a three-dimensional space of $(p_G,p_T,p_{RR})$. In the simplest case, when the threshold surface is a plane and determined by three points $(\pth_G,0,0)$, $(0,\pth_T,0)$, and $(0,0,\pth_{RR})$, it is straightforward to show that
\begin{equation}
    \frac{p}{\pth} = \frac{p_G}{\pth_G} + \frac{p_T}{\pth_T} + \frac{p_{RR}}{\pth_{RR}},
    \label{eq:poverpth}
\end{equation}
where $\pth_G = \pth_G (\eta_G)$ and $\pth_T = \pth_T (\eta_T)$. This formula is in correspondence with Ref.~\cite{google2021exponential}, where $p/\pth \propto \Lambda^{-1}$, and the observed nonlinearities of the threshold surface corroborate with the effects observed in the experiment (see Supplementary Information of \cite{google2021exponential}). In that case Eq.~\eqref{eq:poverpth} implies that the isotropic circuit level noise ($p_G = p_T = p_{RR} = p$) threshold can be recovered as $p_\text{ic}^{-1} = (\pth_G)^{-1} + (\pth_T)^{-1} + (\pth_{RR})^{-1}$, whereas the threshold for the phenomenological noise model ($p_T = p_{RR} = p$) is $p_\text{ph}^{-1} = (\pth_T)^{-1} + (\pth_{RR})^{-1}$. In Tab.~\ref{tab:thresholds} we summarize the parameters of the linearized threshold surface introduced in Eq.~\eqref{eq:poverpth} for the six QEC codes studied in the upcoming sections.

Assuming that a given error configuration is below the error threshold according to Eq.~\eqref{eq:poverpth} one can estimate the number of qubits required for a single logical qubit with practical logical error rate. Let us fix a target logical error rate, i.e., $P_L/w = 10^{-12}$. The required code distance then becomes
\begin{equation}
    d = \left \lceil - \frac{24}{\log_{10}(p/\pth)} \right \rceil,
\end{equation}
implying a total qubit count of $N_\text{tot} = (\nu_q+\nu_a)d^2$ for a single logical qubit, where $\nu_{q(a)}$ is the number of qubits (ancillas) in a unit cell of the QEC code. E.g., for $p/\pth = 0.5$ one needs $N_\text{tot} = 6400(\nu_q+\nu_a) \sim \mathcal{O}(10^4)$ qubits. Less qubits are required as noise decreases. For example, $N_\text{tot} = 576(\nu_q+\nu_a) \sim \mathcal{O}(10^3)$ qubits suffice for $p/\pth =0.1$.

\subsection{Trade-off between measurement time and fidelity}

The maximal-fidelity readout time $\tau^*_R$ in Eq.~\eqref{eq:tRexp} gives the optimal readout error rate if every qubit is measured simultaneously, e.g., at the end of the circuit. However, in quantum error correction data qubits are idling during the measurement of the ancillas, therefore one could expect that it is worthwhile to move away from the maximal readout fidelity point (by reducing the measurement time) in order to improve the idling error rates.

We have seen in Eq.~\eqref{eq:ROerrorrate} how the readout error rate depends on the integration time $\tau_R$. Furthermore, neglecting relaxation for simplicity, the idling error rate during readout and reset reads
\begin{equation}
    p_{T_2} (\tau_R) \approx \frac{1-e^{-(\tau_R+\tau_\text{res})^2/T_2^2}}2,
\end{equation}
where $\tau_
\text{res}$ is the reset timescale. Using $p_{RR} (\tau_R)$ and $p_{T_2} (\tau_R)$ one can derive the readout time that takes the readout-reset and idling error rates furthest away from the threshold surface.

Sticking to our example of a linear threshold surface characterized by $\pth_G$, $\pth_T$ and $\pth_R$, the distance from the threshold surface is given by
\begin{equation}
    \frac p{\pth} = \frac{p_G}{\pth_G} + \frac{p_T (\tau_R)}{\pth_T} + \frac{p_{RR}(\tau_R)}{\pth_{RR}}.
    \label{eq:poverpth}
\end{equation}
In order to reduce the qubit overhead, we need to minimize this ratio with respect to $\tau_R$. Assuming that $\tau_{R}\ll T_2, T_{1\text{RO}}$ the readout time that is optimal for QEC applications becomes
\begin{equation}
    \tau^*_{\text{QEC}} = \tau_R^* - \tau_\text{min} \log \left( 1+  \frac{\pth_R}{\pth_T}\frac{(\tau^*_{\text{QEC}}+ \tau_\text{res})T_{1R}}{(1-2p_\text{res})T_2^2} \right).
    \label{eq:tauQECoptimal}
\end{equation}
This equation can easily be solved recursively, starting from $\tau^*_{\text{QEC}} = \tau_R^*$ on the right-hand side. As we will see later on the example of the surface code, it is even possible that the maximal fidelity readout time falls outside the threshold surface, while $\tau^*_{\text{QEC}}$ is deep inside it.

\begin{figure*}
    \includegraphics[width = \textwidth]{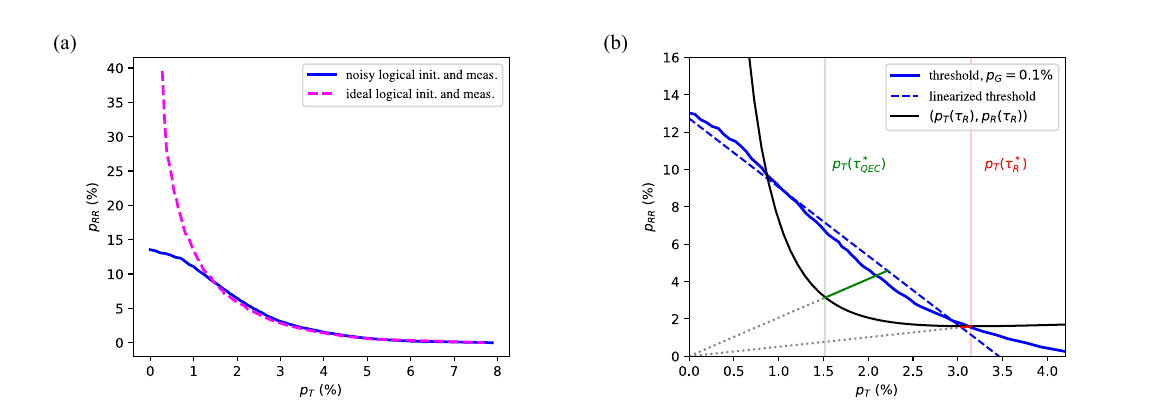}
    \caption{(a) Threshold curves for ideal logical state preparation (magenta) and noisy logical state preparation in a quantum memory simulation. At low idling error rates the ideal logical preparation and measurement significantly overestimates the threshold. Threshold values are obtained from code distances $d\in \{11,13,15,17\}$ and $T=d$ stabilizer cycles. (b) Readout error  rate \eqref{eq:ROerrorrate} for varying integration (solid black) time, threshold curve (solid blue) for $p_G = 0.1\%$ and the linearized threshold curve (dashed blue). The integration time corresponding to the maximum fidelity ($\tau_R^*$) falls outside the threshold, while setting the integration time for the QEC-optimum ($\tau^*_{\text{QEC}}$) remains deep inside the threshold. Numbers used in this exemplary plot are $\tau_\text{min} = 1.5\mu$s, $T_{1R} = 1$ms, $\tau_\text{res} = 3\mu$s, $p_\text{res} = 0.5\%$, and $T_2 = 50\mu$s.}
    \label{fig:optimal_readout}
\end{figure*}

\subsection{Surface code with circuit-level noise}

Let us consider Kitaev's surface code with rotated boundaries \cite{bombin2007optimal}. In this QEC code, $d^2$ data qubits are arranged in a square grid and the group of stabilizer operators contain $(d-1)^2$ plaquettes with products of Pauli-X and Pauli-Z operators acting on four nearest-neighbor qubits. At the boundaries of the lattice one needs $2d-2$ stabilizers with support on only two qubits each yielding $d^2-1$ constraints for the $d^2$ degrees of freedom. Measuring all the stabilizers is then equivalent to completing four- and two-body parity measurements on neighbouring qubits. In what follows, we consider the standard circuit representation of the surface code and consider individual errors during a quantum memory experiment that can lead to non-local error chains and show a way to neutralize them. Afterwards, a possible experimental protocol of initializing and reading out a logical qubit is discussed, assuming noisy ingredients throughout the entire process.

Provided that every qubit can be read out individually, the four-body stabilizer measurements of the surface code can be achieved by adding ancilla qubits at the center of every plaquette leading to denser square grid of qubits \cite{dennis2002topological}. Ancillas are reset to the state $\ket 0$ before every stabilizer round and therefore do not add degrees of freedom to the system. The Z-plaquette measurements are then implemented via four controlled-not (CX) gates controlled on corresponding data qubits and targeted on the ancilla. The controlled gates will flip the state of the ancilla an even or an odd number of times depending on the parity of the four data qubits. Measuring the ancilla then yields the eigenvalue of the Z-stabilizer.

Assuming any gate in the stabilizer measurement circuit can induce an error in the corresponding part of the circuit, errors can propagate from the ancilla qubit to the data qubits via the CX gates (e.g., X errors spread from the control to the target, and Z errors from the target to the control). The worst-case-scenario errors --called {\it hook errors} in the literature-- are the X (Z) errors in the X-stabilizer (Z-stabilizer) circuits that occurs before the third CX gate, because this error will spread to two of the data qubits~\cite{dennis2002topological}. Four (three) data qubit errors make up a full stabilizer (a full stabilizer and a single error), that is less harmful than two data-qubit errors. In the rotated surface code, the X (Z) logical operator is a Pauli string that acts on a column (row) of qubits. Depending on the schedule of CX gates, hook errors can be a distance-2 sub-string of a logical operator, meaning that half as many of them is required to induce an undetectable logical error, i.e., reducing the effective code distance to $\lceil d/2 \rceil$. With careful scheduling, however, it can be ensured that no single error event in the circuit can induce errors that reduce the code distance in the surface code~\cite{dennis2002topological}.

\subsubsection{Noisy logical initialization and readout}

A quantum memory experiments consist of the initialization of the logical qubit in the Z (X) basis followed by several rounds of stabilizer measurements and the final readout of the logical qubit on the same basis. Afterwards a decoder is used to infer from the syndrome data if a logical X (Z) error happened. Comparing the initialized logical eigenvalue to the final corrected logical readout of the simulation allows one to calculate logical error rates for different physical error rates and code distances to determine the error threshold.

The initialization of the logical qubit is often carried out assuming noise-free initial and final stabilizer measurements as well as perfect initialization and measurement of the data qubits. Here we briefly review this protocol to show that it leads to an unphysical solution for the threshold surface, and compare this result with a fault-tolerant protocol we used in our work, where every qubit and quantum gate is subject to the noise model introduced in Sec.~\ref{sec:noisemodel}.

In the case of the ideal logical initialization and readout protocols one prepares every qubit in the $\ket{0}$ ($\ket +$) state which is an eigenstate of the logical $Z$ ($X$) operator. Performing the first round of stabilizer measurements, the $X$ ($Z$) stabilizer outcomes will be random even in the absence of errors, since the physical qubit resets did not prepare a surface code state. Since stabilizers commute with the logical operator, the logical eigenvalue is still intact after the first round of stabilizers, but the system is projected into a subspace of some given $X$ ($Z$) stabilizer eigenvalues. From the second round, every stabilizer measurement would return the same syndrome as in the first one, so one can start inserting errors and the decoder will have enough information to deal with the logical correction. The final measurement of the logical follows a similar logic: performing a noise-free round of stabilizer measurements before reading out the physical qubits that reveal the encoded logical eigenvalue. It is easy to see that the logical eigenvalue cannot be corrupted in this protocol if only readout or reset errors happen during the noisy stabilizer rounds. 

In a real experiment, however, the initialization and final readout are noisy processes as well. If the errors are strongly polarized towards readout errors, the logical readout will be heavily affected by errors, implying a finite threshold error value against readout errors. The preparation of the initial state can be performed fault-tolerantly with noisy operations only as well. If a code contains pure Pauli-X and pure Pauli-Z stabilizers and logical operators, only the Z-stabilizer (X-stabilizer) outcomes are needed to determine whether a logical Z (X) error happened, since X (Z) errors do not affect the X-stabilizers (Z-stabilizers). We can, therefore, prepare every qubit in the $\ket 0$ ($\ket+$) state initially and read out every qubit on the Z (X) basis in the end. The initial data qubit resets ensure that all Z (X) stabilizer eigenvalues are known before the first round, and final measurements can be used to infer the relevant syndromes after the last round of stabilizer measurements, thereby allowing for errors to be injected at any point in the process \cite{fowler2018low}.

\section{Surface code with spin qubits}
\label{sec:surface}

\begin{figure*}
    \includegraphics[width=\textwidth]{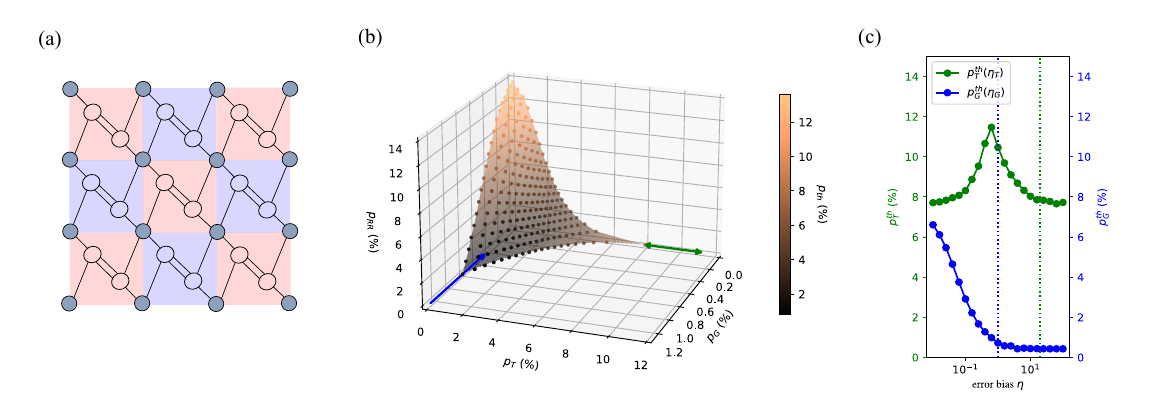}
    \caption{(a) Qubit layout and connectivity map for a bulk section of the rotated surface code (b) error-threshold surface in the parameter space of gate- ($p_G$), idling- ($p_T$), and readout error rates ($p_R$) with biases set to $\eta_T = 20$ and $\eta_G =1$. Each point is calculated from code distances $d\in \{11,13,15,17\}$ and $T=d$ stabilizer cycles. Green and blue arrows indicate the change of $(p_G,0,0)$ and $(0,p_T,0)$ corner points of the surface for a wide range of noise biases. (c) Dependence of $\pth_G$ on the gate error bias $\eta_G$ and $\pth_T$ on the idling error bias $\eta_T$.}
    \label{fig:threshold_RSC}
\end{figure*}

In the previous section we have seen that the surface code can be conveniently embedded in a square lattice of qubits. However, since spin-qubit readout requires two qubits, the square-grid qubit layout needs to be modified. An example for such a modified lattice was provided by Ref.~\cite{veldhorst2017silicon} and shown in Fig.~\ref{fig:SurfCodeLayoutCircuit}(a). Data qubits are connected to four neighbouring ancillas via two-qubit gates but do not participate in a readout pair. Ancillas, on the other hand, always come in pairs where they can be swapped, reset, or read out in a single step.

A pair of ancillas can be initialized directly in the $\ket{00}$ state with a fidelity $\mathcal F_\text{init}$ and their Z-parity can be read out in a partially destructive process \cite{seedhouse2021pauli,niegemann2022parity} we denote as a $ZZ^*$ measurement box. Keeping one of the ancillas in the $\ket 0$ state as a reference, the Z-stabilizer measurement circuits can be realized using a single SWAP gate as shown for the upper ancilla pair on Fig.~\ref{fig:SurfCodeLayoutCircuit}(b). The X-stabilizer measurements, on the other hand, require in total three SWAPs (see the lower ancilla pair of Fig.~\ref{fig:SurfCodeLayoutCircuit}(b)). Without the additional SWAP gates, both X and Z plaquettes are bound to have the same (or equivalent) CX schedules. Consequently, X-error pairs and Z-error pairs are injected in the same direction, reducing the code distance for either of the logical operators regardless of the choice of boundary conditions. Alternative stabilizer measurement schemes, trying to leverage the second ancilla qubit, can be found in App.~\ref{smsec_stab_meas}.

If we restrict our attention to memory experiments, it is tempting to propose a qubit connectivity where the ancilla pairs are placed such that the the number of SWAP gates enforced by the layout is minimized. At the same time, moving towards fault-tolerant quantum computing with multiple logical qubits, such a hard-wiring of the CX schedule in the physical qubit layout is not possible. E.g., twist defects require an on-demand change in the checkerboard pattern of plaquette operators \cite{brown2017poking}.

Fault-tolerant logical-state preparation and readout requires data qubits to be initialized and read out. This can be done by adding a single pair of ancillas to the surface code lattice, allowing one to assign one ancilla pair to every code qubit. Initialization is done by resetting the ancillas, swapping the data qubit with the ancilla it is connected to, and performing a second reset on the ancillas. Similarly the final measurement can be carried out using a single layer of SWAP gates. Consequently, data qubit measurement does not require additional hardware. Further details about the CX schedule, logical initialization and measurement can be found in App.~\ref{smsec_surface}.

Calculating the logical failure rate as a function of $p = (p_T^2 + p_{RR}^2)^{1/2}$ for different code distances, the intersection of failure rates yields the threshold for a given $p_{RR}/p_T$. Repeating this for several distinct ratios one obtains a threshold curve in the parameter space of $(0,p_T,p_R)$. In addition, it is important to consider the logical qubit initialized both along the $X$ and $Z$ axes (taking the smaller threshold), since the biased idling noise can strongly favour one type of logical operator. On Fig.~\ref{fig:optimal_readout}(a) we compare the threshold curves with the logical initialization and readout performed using ideal stabilizer circuits and the fault-tolerant protocol for $p_G = 0$. As expected, only the fault-tolerant protocol leads to finite threshold for readout errors only showcasing the importance of the logical initialization protocol in memory-experiment simulations.

Using the linear approximation of the threshold curve, we show that in certain cases the maximal readout fidelity is far from the optimal choice for error correction. For the parameter values on Fig.~\ref{fig:optimal_readout}(b) the maximum fidelity ($\mathcal F_R \approx 99\%$) is achieved at $\tau_R^* = 9.8\mu$s, whereas the optimal choice is $\tau^*_\text{QEC} = 5.8\mu$s for $T_2 = 50\mu$s. Using the optimal readout time, we are deep inside the fault-tolerant regime, while the integration time for the maximal readout fidelity leaves so much time for decoherence on the data qubits that the rate of success decreases with increasing system size.

\begin{figure*}
    \includegraphics[width=\textwidth]{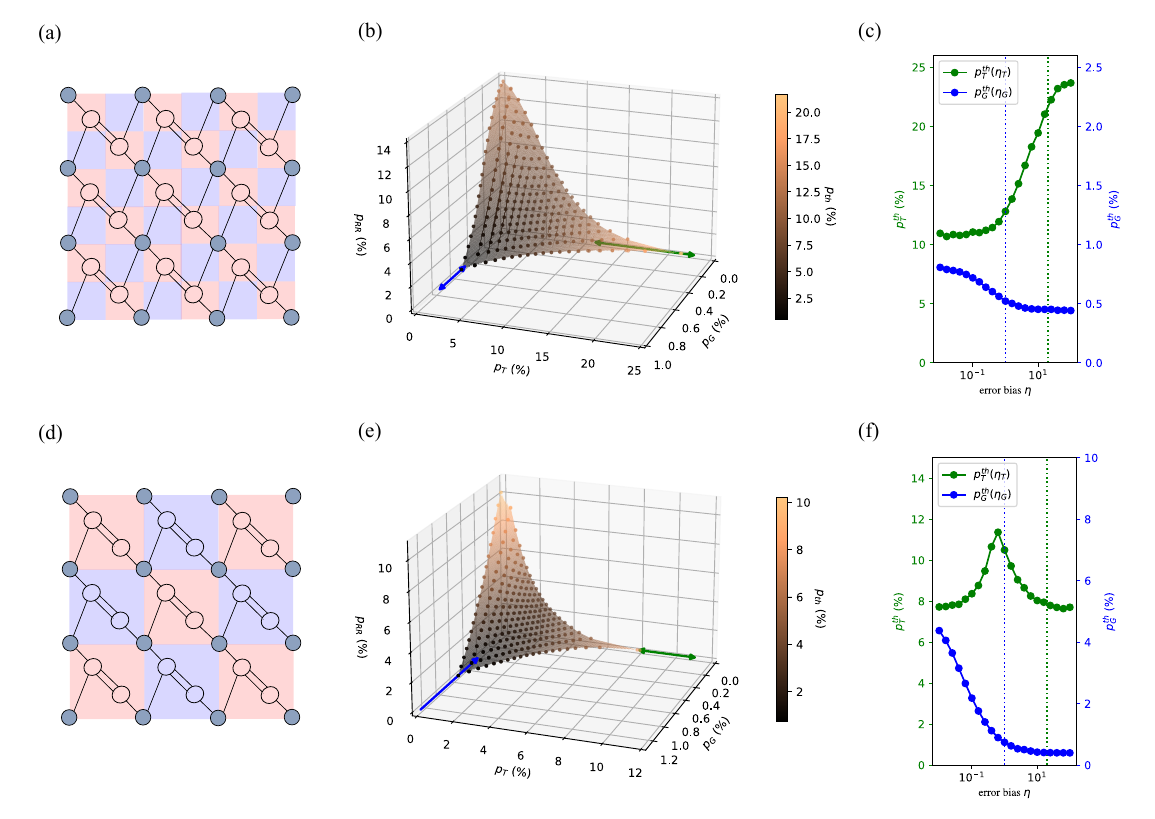}
    \caption{(a) Qubit layout and connectivity map for a bulk section of the XZZX code (b) threshold surface for $\eta_T = 20$ and $\eta_G =1$, (c) Dependence of $\pth_G$ on the gate error bias $\eta_G$ and $\pth_T$ on the idling error bias $\eta_T$ for XZZX code. Each point is calculated from code distances $d\in \{11,13,15,17\}$ and $T=d$ stabilizer cycles. (d)-(f) Similarly, qubit layout, threshold surface, and bias dependencies of the 3-CX surface code.}
    \label{fig:threshold_XZZX_3CX}
\end{figure*}

The analysis so far considered an optimistic fixed value for the gate errors. Although gate errors make substantial contribution to the overall performance of error correction. In order to gain a more complete picture of the threshold surface we calculated the threshold surface in the three-dimensional parameter space of $(p_G,p_T,p_R)$ for error biases $\eta_G = 1$ and $\eta_T=20$. The results are presented on Fig.~\ref{fig:threshold_RSC}(b). The error threshold for a gate-error-dominated scenario [i.e., $\mathbf{p} \approx (p_G,0,0)$] is substantially lower than the threshold values of the opposite limit. Even though the threshold surface does not have the shape of a plane, we may fit a plane to the data points using the three threshold parameters from Eq.~\eqref{eq:poverpth}, to allow for a simple quantitative comparison between QEC codes as well as a useful proxy for resource estimation. The quality of the fit can be described by the mean distance between the surface and the fitted plane, which can be used to estimate the error of the linear approximation (see Tab. \ref{tab:thresholds}). Due to the monotonicity of the threshold surface, one expects the strongest bias dependencies in the corners, i.e., for $(0,\pth_T(\eta_T),0)$ and $(\pth_G(\eta_G),0,0)$. Therefore we focused our error-bias analysis on these two points.

From Fig.~\ref{fig:threshold_RSC}(c) it is apparent that the idling threshold $\pth_T$ is peaked at $\eta_T = 1/2$, that corresponds to depolarizing noise. This can be understood from the fact that for $\eta_T \ll 1$, $X$ and $Y$ errors occur with a probability $p_{X-Y} = p$, both contributing to logical $X$ errors. Similarly, for the $Z$-biased case ($\eta_T \gg 1$) the probability-$p$ errors contribute to logical-$Z$ errors. On the other hand for depolarizing noise, only two of the three Pauli-s affect a logical with a joint probability $2p/3$. Indeed we observe a roughly $1.5\times$ increase in the error threshold. We note that noise-bias-tailored decoders could take advantage of the peculiar syndrome pattern of biased noise leading to potentially higher thresholds~\cite{tuckett2020fault}, but such an advantage has not been demonstrated yet for circuit-level error models.

Furthermore, there is a significant increase in gate threshold $\pth_G$ for errors biased towards single-qubit gate errors on Fig.~\ref{fig:threshold_RSC}(c). This can be understood from the stabilizer measurement circuit in Fig.~\ref{fig:SurfCodeLayoutCircuit}(b). Only X-stabilizer circuits include single-qubit gates which only contribute to faulty syndromes. Faulty syndromes being equivalent to readout errors have a high threshold.

\section{Error correction beyond Kitaev's surface code}
\label{sec:comparison}

Now we turn our discussion towards other members of the surface code family. In the recent years several new candidates appeared that challenge the surface code in terms of error threshold \cite{ataides2021xzzx}, required connectivity \cite{Hastings2021dynamically,kesselring2022anyon, mcewen2023relaxing}, and prospects for fault-tolerant logical gates~\cite{Wootton_2015,srivastava2022xyz}. 

Among the codes considered here, there are codes that can be obtained from the Calderbank-Shor-Steane (CSS) construction~\cite{calderbank1996good} using only pure $X$ and $Z$ type of Pauli operators, while some other codes include mixed stabilizers and logicals. As we will show, candidates from the latter class tend to perform better against biased idling errors than the CSS counterparts.

Furthermore, we analysed two Floquet codes, where the stabilizer operators change periodically from one stabililzer round to the next~\cite{Hastings2021dynamically,kesselring2022anyon}. Such a scheme facilitates the measurement of six-body stabilizer operators with very low connectivity (i.e., $2\tfrac 1 4$ two-qubit links per qubit on average).

\subsection{The closest relatives: the XZZX and the 3-CX surface code}

The XZZX code can be simply derived from the rotated surface code, by exchanging $X$ and $Z$ Pauli operators along one of the diagonals for every plaquette such that every plaquette operator, from left to right and top to bottom consist of Pauli operators $X$, $Z$, $Z$, and $X$~\cite{ataides2021xzzx}. Logical operators need analogous adjustments to maintain the necessary commutation relations. See App.~\ref{smsec_surface} for a more detailed example. The XZZX code requires the same connectivity as the rotated surface code (see Fig.~\ref{fig:threshold_XZZX_3CX}(a)), but the local basis transformations necessitate extra single-qubit gates on the data qubits which increase the circuit depth and the error budget of gates. On the other hand the XZZX code brings significant improvement in terms of idling threshold compared to the surface code as can be seen on the scale of the $p_T$ axis on Fig.~\ref{fig:threshold_XZZX_3CX}(b). 

From the gate-bias dependence of the respective corner point (shown on Fig.~\ref{fig:threshold_XZZX_3CX}(c)) it is clear that the XZZX code can present a real advantage over the surface code only if two-qubit gate errors are more likely. The most remarkable property of the threshold surface is the dependence of the idling threshold on $\eta_T$. Since the logicals are not pure Pauli $X$ and $Z$ operators as for the surface code, the maximum is not achieved for depolarizing noise $\eta_T = 1/2$ but in the limit where $\eta_T \gg 1$, which we believe to be the relevant limit for spin qubits (i.e., $T_2 \ll T_1$).

The second candidate, the 3-CX surface code, presented recently by Ref.~\cite{mcewen2023relaxing} measures surface-code stabilizers in a two-round stabilizer measurement cycle such that stabilizers are measured once per round, but the state of the data qubits only returns to the original state in every second round. In this peculiar measurement sequence, only three of the four connections are used for every data qubit (see Fig.~\ref{fig:threshold_XZZX_3CX}(d)), reducing the required connectivity to an effective hexagonal grid. In App.~\ref{smsec:3CXdetregs} we briefly review the detecting region picture developed by Ref.~\cite{mcewen2023relaxing} and provide the stabilizer measurement circuit for the $d=2$ instance of the code.

The measurement sequence of the 3-CX surface code requires a modified spatial boundary compared to the rotated surface code \cite{mcewen2023relaxing} that we discuss in App.~\ref{smsec:3CXdetregs}. Moreover the fault-tolerant logical readout suggested in Ref.~\cite{mcewen2023relaxing} involves simultaneous measurement of all data and ancilla qubits. Although we only have the hardware to read out half of the qubits in our spin qubit lattice. In order to overcome this issue, we developed an improved final readout for which the ancilla-only readout pairs of Fig.~\ref{fig:SurfCodeLayoutCircuit}(b) are sufficient.

In QEC codes, reduced connectivity often comes at the expense of a significantly reduced threshold~\cite{jones2018logical}, however, Figs.~\ref{fig:threshold_XZZX_3CX}(e)-(f) show, in correspondence with the findings of Ref.~\cite{mcewen2023relaxing}, that no significant compromise was made by adapting the stabilizer measurement sequence of the surface code to a lower-connectivity qubit lattice.

\begin{figure*}
    \includegraphics[width=\textwidth]{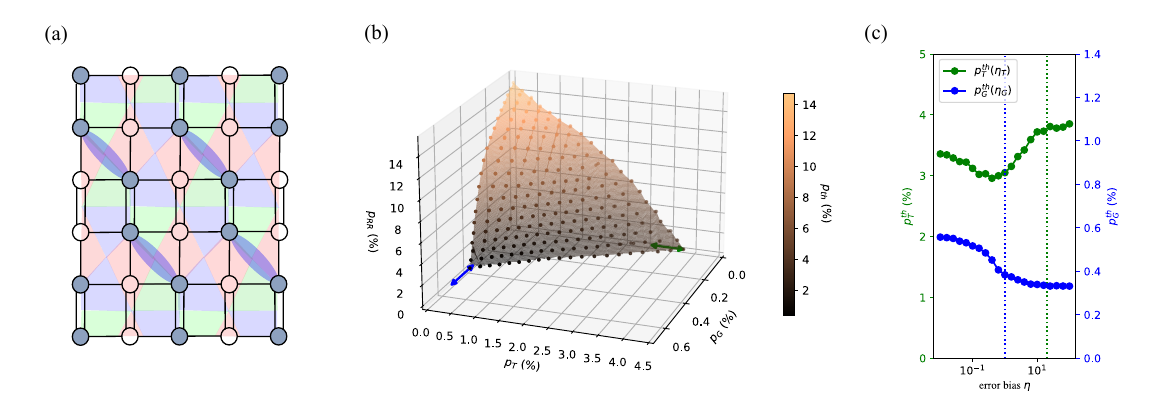}
    \caption{(a) Qubit layout and connectivity map for a bulk section of the XYZ$^2$ matching code. Plaquette stabilizers are $XYZXYZ$ Pauli products acting on the the six data qubits around each ancilla pair and $ZZ$ link stabilizers are shown with thick blue lines. Data qubit readout system is required for the final logical readout (see App.~\ref{smsec:xyz2} for further details). (b)-(c) Threshold surface and bias dependencies of the XYZ$^2$ matching code for $\eta_T = 20$ and $\eta_G =1$. Each point is calculated from code distances $d\in \{11,13,15,17\}$ and $T=d$ stabilizer cycles.}
    \label{fig:threshold_XYZ2}
\end{figure*}

\subsection{XYZ$^2$ matching code}

Hexagonal matching codes are special class of $D(\mathbb Z_2)$ anyon models where hexagonal plaquette stabilizers are bicolorable and host different anyon species with the required braiding statistics. Fermionic quasiparticles, combined from two anyons of different species, are confined to string stabilizers that connect the same coloured plaquettes~\cite{Wootton_2015}. In larger-scale lattices the confined-fermion property of matching codes allows for twist-defect-based logical operators without introducing five-body stabilizers as for the surface code.

The XYZ$^2$ code is a variant of the hexagonal matching codes where the string stabilizers are parallel links on the hexagonal lattice \cite{srivastava2022xyz}. Suitable boundary conditions of the code can be found from the concatenation of a two-qubit repetition code (stabilized by $ZZ$) and the XZZX. In total we get six-body stabilizer operators with $X$, $Y$, $Z$, $X$, $Y$, and $Z$ Paulis in clock-wise direction around the hexagonal plaquettes, and $ZZ$ link along the $XY$ edges of the hexagonal plaquettes.

Stabilizer measurements in a spin qubit architecture can be achieved in two rounds such that there is a pair of qubits in the face of each hexagon as shown in Fig.~\ref{fig:threshold_XYZ2}(a). The scheduling of CX gates need to follow similar considerations to the surface code case in order to prevent the injection of hook errors. In App.~\ref{smsec:xyz2} we show that there is such a choice of CX schedule for the XYZ$^2$ code.

The price to be paid for the dense encoding, is that there are less ancilla pairs than stabilizers. Meaning that for the final readout of the logical and inference of relevant stabilizers, the code qubits also need to be part of readout-pairs. However, in terms of connectivity the layout still remains a regular grid of qubits  (as discussed in detail in App.~\ref{smsec:xyz2}) realizable e.g., with $2\times N$ arrays and long-range links. Since only the spin-blockade requires immediate proximity, parallel rows of readout pairs can be accommodated by bi-linear arrays and connected via long-range couplers.

The threshold surface of the XYZ$^2$ code is closer to a plane than that of the previous surface code variants (see Fig.~\ref{fig:threshold_XYZ2}(b)). Interestingly, the readout threshold remains comparable to that of the surface code even though the stabilizers are read out in separate measurement rounds. Not being a CSS code, the idling threshold is not peaked around the depolarizing limit and due to different $X$ and $Y$ basis conversions in the plaquette measurements single qubit gate errors contribute significantly to the error budget, i.e., error threshold biased towards single-qubit errors is not significantly higher than the ones biased towards two-qubit gate errors. These results are presented in Fig.~\ref{fig:threshold_XYZ2}(c).

\subsection{Floquet codes}

A new type of stabilizer codes has been proposed recently by Hastings and Haah \cite{Hastings2021dynamically}, which do not have a static stabilizer group and logical operators, but they change periodically over six rounds. Their proposal was based on Kitaev's honeycomb model (which as a static stabilizer code has no logical qubits). An even more recent example is the Floquet color code. This is a CSS code that can be obtained from a color code using the anyon condensation picture of Ref.~\cite{kesselring2022anyon}. Since the results for both codes are quantitatively very similar, we will present only those for the latter in the main text. We defer a full comparison of the two to App.~\ref{smsec:floquet} (see Fig.~\ref{smfig_honeycomb_vs color_floquet}).

A brief summary of the anyon condensation picture can be found in App.~\ref{smsec:floquet}, here we only consider the stabilizer group after different measurement rounds. The code can be realized on a hexagonal lattice of data qubits placing an ancilla pair to every edge of the lattice. From a an architecture standpoint Floquet codes require a very sparse connectivity, only $2\tfrac 1 4$ links per data qubit leaving valuable space for wire routing and other elements of the control circuitry.

\begin{figure*}
    \includegraphics[width=\textwidth]{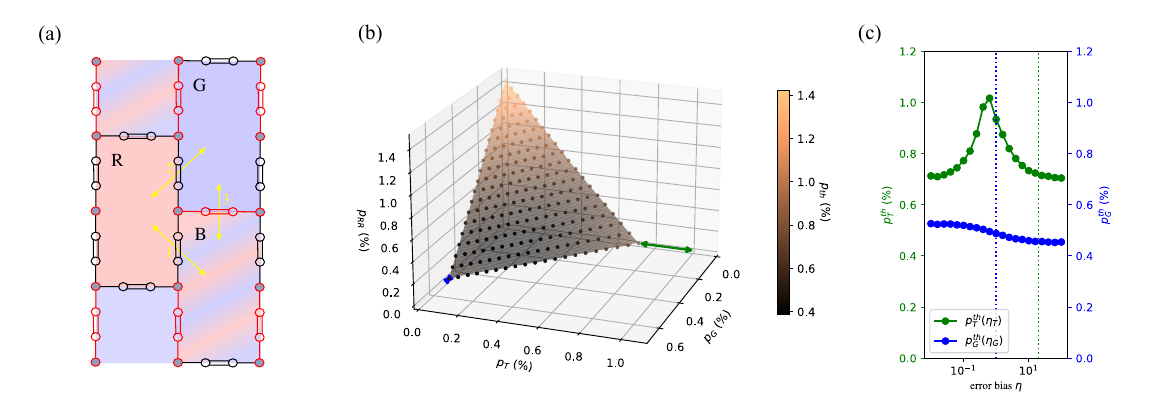}
    \caption{Qubit layout and connectivity map for a bulk section of the Floquet color code. After measuring the red-highlighted XX-links, additional plaquette stabilizers are 6-qubit Pauli-X stabilizers on R-labelled plaquettes, Pauli-Z stabilizers on G-labelled plaquettes and both X- and Z-stabilizers on B-labelled plaquettes (notation is coming from the parent color code). In each time step a set of links is measured such that the six-qubit stabilizer types are exchanges between two plaquette types, indicated by yellow arrows. (b)-(c) Threshold surface and bias dependencies of the Floquet color code for $\eta_T = 20$ and $\eta_G =1$. Each point is calculated from code distances $d\in \{10,12,14,16\}$ and $T=d/2$ stabilizer cycles (six rounds per cycle).}
    \label{fig:threshold_FCC}
\end{figure*}

The hexagonal lattice is tricolorable such that every hexagon has neighbours with a different color. We label them as R, G, and B. In every round one measures a set of links that are matching same-labelled hexagons on some basis. E.g., starting from the stabilizer group depicted in Fig.~\ref{fig:threshold_FCC} --where R plaquettes are hosting only a pure Pauli-X operator, G plaquettes only Pauli-Z, and B plaquettes both types-- one measures green links (matching G plaquettes) on the Z basis to effectively exchange the roles of R and B while leaving G intact. Such exchange of roles can be performed in a cyclic manner using differently colored X and Z links only to arrive in the same state after six rounds. We point out that leaving one species of plaquettes with the same stabilizers in each round is crucial for the preservation of the logical information.

In the SWAP-based syndrome measurement scheme we have used so far, one would need two CX gates and a SWAP in-between to read out the link stabilizers. However, another stabilizer measurement protocol can also be employed that avoids using SWAP gates thereby reducing the gate-error budget of the noise model (see also App.~\ref{smsec_stab_meas}). If the ancilla state is prepared in the $(\ket{00}+\ket{11})/\sqrt 2$ Bell-state, CX-s can target the two ancillas independently, such that the measurement of the ancilla-pair remains deterministic. For the results presented in Fig.~\ref{fig:threshold_FCC}(b)-(c) we employed this Bell-state protocol.

The threshold surface of the Floquet color code is very well approximated by a plane (see Fig.~\ref{fig:threshold_FCC}(b)). This property could be attributed to {\it (i)} the fact that during a link measurement only a single data qubit error can be injected, {\it (ii)} the lack of fault-tolerant logical initialization and readout. The bias dependencies on Fig.~\ref{fig:threshold_FCC}(c) of the gate and idling thresholds show the expected behaviour for CSS codes. 

Even though proposals exist for the spatial boundaries of both types of Floquet codes \cite{kesselring2022anyon, Haah2022boundarieshoneycomb}, here we only studied these codes with toric spatial boundary conditions for simplicity. Likewise, we omitted the problem of fault-tolerant initialization and readout of the logical qubit since the multi-round stabilizer measurement protocol gives a finite (and relatively low) readout-reset threshold.

\subsection{Comparison of QEC codes}

As discussed previously, a linearized threshold surface could, through simple formulas, yield valuable insight about the device-specific threshold and the optimal measurement time for mid-circuit measurements. Here we focus on the corner points of the linearized threshold surface obtained by fitting a plane to the simulated data. The characterization of the surface with only three numbers aids simple estimates for present and future experiments determining optimal readout parameters for error correction as well as estimating the device-specific error threshold and qubit overhead for different spin qubit architectures from high- to low connectivity.

Our results are summarized in Tab.~\ref{tab:thresholds}. The three threshold parameters allow us to quantify some trends for spin-qubit devices expected from the QEC literature. E.g., error thresholds rapidly decrease by reducing the connectivity, the limiting case being linear qubit chains (connectivity $=2$) that are shown to have thresholds $\pth=10^{-5}$-$10^{-4}$ for multiple linear QEC codes~\cite{jones2018logical}. Further, we see that gate thresholds are up to an order of magnitude smaller than the other two threshold parameters in agreement with the results on the surface code where thresholds for circuit-level noise are significantly lower than that of simplified noise models.

An example how our results may be utilized is by considering a hypothetical scenario, when good single- and two-qubit gates are available ($p_G \sim 0.1\%$), but the readout is very limited, e.g., during the minimum integration time of the ancilla readout, idling errors on data qubits reach up to a few percent probability. This would exclude the surface code and many more from our comparison in Tab.~\ref{tab:thresholds} but the XZZX code, featuring high idling threshold parameter, can still be utilized. Having identified the qubit layout to be built, the integration time can be set according to Eq.~\eqref{eq:tauQECoptimal} in order to reach the best error correcting performance.

The QEC codes studied are all defined in a two-dimensional qubit layout with local two-qubit connections. This implies that the number of data qubits $n$ scales with the square of the code distance, i.e., $n = \nu_q d^2$ \cite{bravyi2010tradeoffs}. The total number of qubits required including both data and ancilla qubits $N_\text{tot} = (\nu_q+\nu_a)d^2$ can also be accessed from Tab.~\ref{tab:thresholds}. The lowest qubit overhead corresponds to the closest relatives of the rotated surface code coming at $N_\text{tot} = 3d^2$. Exploiting the double ancillas required by the spin-blockade readout, the XYZ$^2$ code uses only $4d^2$ qubits to measure six-body stabilizer operators as opposed to $6d^2$ of the Floquet codes.

Finally we make some quantitative comparisons to values found in the literature for the different QEC codes.
disregarding the finite idling bias, for the rotated surface code in the presence of isotropic circuit level noise we get $p_\text{ic} = 0.65\%$ and for the phenomenological noise model $p_\text{ph} = 3.1\%$ is obtained, which are in line with the expectations \cite{raussendorf2007topological,tuckett2020fault}. For the XZZX code we obtain $p_\text{ic} = 0.35\%$ and $p_\text{ph} = 6.9\%$, where the latter is in good agreement with Ref.~\cite{ataides2021xzzx}. Some deviations are to be expected for these threshold values due to the differences in the stabilizer measurement protocol, e.g., the use of pair-wise ancilla readout and the additional SWAP gates.

\section{Discussion}
\label{sec:discussion}

Some important characteristic properties of spin qubit architectures are taken into account in our model and the threshold surface analysis allows one to tailor the results to system-specific features. However, some assumptions on the noise model need further improvement for more accurate device-level threshold estimates. In particular, gate errors as well as readout errors were modelled here as depolarizing noise, which may have tendencies to be more like either bit-flip or phase errors in a specific hardware. Also, correlated noise due to cross-talk can be taken into account within the Pauli-error model, but it requires a high-level characterization of the envisioned quantum processor. Finally, some types of errors like coherent errors (systematic rotation of every qubit with the same angle) are not captured in our model and need to be analyzed with separate methods~\cite{bravyi2018correcting}.

Restricting the focus regarding the qubit layout and the expected noise model would help to develop noise-tailored decoders~\cite{tuckett2020fault} with improved performance compared to the one utilized in our work. Fault-tolerant quantum computation in scalable architectures will also require real-time decoding, the performance of which is likely to be compromised by the available time budget. A detailed study is required to identify the optimal decoding strategy which provides the highest thresholds for the given noise model and available decoding time.

Since the Bell-state protocol introduced for the link measurements of Floquet codes leads to a reduced number of SWAP gates, it is tempting to try a similar strategy for all other codes considered. However, note that some of the SWAP gates on the ancillas are necessitated by the specific CX schedule to avoid errors that would reduce the code distance. These SWAP gates are still required in the latter protocol. Furthermore, in the Bell-state protocol, it is the reset errors of the ancillas that would propagate to the data qubits as opposed to gate errors in the SWAP-based method, reducing the readout-reset threshold in the former case.

Our findings can support the development of new QEC codes tailored to spin-qubit architectures. One possibility is to combine XZZX stabilizers with the measurement sequence of the 3-CX surface code to obtain a lower connectivity error correction code with high idling threshold error rate at strongly biased noise. Floquet codes also provide promising prospects for future codes with low connectivity.

Finally, the best error correction code will be the one that provides the lowest qubit overhead en route to fault-tolerant quantum computing. Beyond the quantum memory experiments considered here, fault-tolerant logical gate schemes using twist defects or lattice surgery~\cite{brown2017poking} also need to be revisited under concrete noise models to find the lowest space-time overhead for a given device design. 

For the Clifford-circuit simulations we used stim~\cite{Gidney2021stim}. For each code distance and physical error rate we took up to $300\,000$ shots unless $30\,000$ logical failures are encountered before that. For the decoding we used pymatching~\cite{higgott2022pymatching}. All the scripts used for the threshold calculations as well as the plotted data are available at~\cite{hetenyi2023github}.

\section{Conclusion}
\label{sec:conclusion}

Taking into account the common features of spin-qubit platforms we have derived an error model accounting for different error rates for gate and readout errors as well as decoherence during mid-circuit measurements. This helped us to quantify the trade-off between fast and accurate qubit measurements for error correction applications. We derived a formula for the optimal integration time that can be used for the calibration of the qubit readout given the noise parameters of the device and QEC code of interest.

Considering state-of-the-art error correction codes that are compatible with locally connected 2D architectures, we proposed four different qubit layouts required for quantum memory experiments which offer the spin qubit community different options to find balance between experimental feasibility and low-overhead execution of QEC codes. Furthermore we analyzed the threshold surface in a multi-dimensional parameter space facilitating back-of-the-envelope estimates for the error threshold and qubit overhead involving on the gate and readout fidelities as well as the decoherence rates for a given experimental setup.

\section{Acknowledgements}

We thank B. Srivastava and the spin qubit team at IBM Research - Zurich the useful discussions, B.J. Brown for providing access to the script on the Floquet color code, and J. Asb\'oth for pointing out some relevant references.

The authors acknowledge support from the NCCR SPIN, a National Centre of Competence in Research, funded by the Swiss National Science Foundation (grant number 51NF40-180604).

\appendix

\section{Spin qubit readout and stabilizer measurements}
\label{smsec_stab_meas}

\begin{figure}
\includegraphics[width=0.49\textwidth]{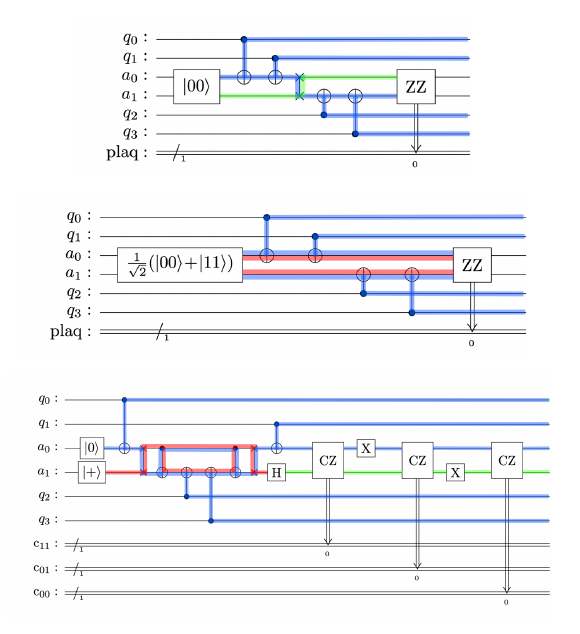}
\caption{Alternative stabilizer measurement protocols with two ancilla qubits and pairwise readout. Top: SWAP-based protocol with destructive parity measurement (ZZ) which is replaceable by single-state discrimination (CZ). Middle: Bell-state measurement protocol reducing the SWAP gate count. Bottom: second ancilla is utilized as a flag qubit requiring subsequent measurements with single-state discrimination. Overlay: detecting regions expanding from the reset-measurement pair, blue and green denote Z type regions, red denotes an X type region.}
\label{smfig_2ancilla_stabilizer}
\end{figure}

In order to discuss the prospects of stabilizer measurements with spin qubits we turn to the example of the Z-plaquettes of the surface code and present three options for performing stabilizer measurements using an ancilla pair instead of the conventional single ancilla that can be read out directly. Excluding readout protocols that require a lead coupled to the quantum dot hosting the qubit, we focus on readout via Pauli spin blockade. 

In the Pauli spin blockade, the two-qubit subspace corresponds to the four low-energy two-particle eigenstates of a symmetric double quantum dot with one particle residing in each dot. When the energy of one of the dots is increased, a particle tunneling from the higher-energy dot into the lower-energy one is only allowed into the spin singlet state (triplets involve a higher-energy orbital due to the exclusion principle). This protocol converts one of the states into a doubly-occupied charge state that is measurable via a charge sensor. The rest of the states can only end up in the measurable charge-configuration by relaxation \cite{vigneau2022probing}.

The above-described effect can be utilized for readout in two different regimes: {\it (i)} destructive parity measurement, or {\it (ii)} single-state discrimination~\cite{seedhouse2021pauli,niegemann2022parity}. 
In the destructive parity measurement protocol, one exploits the fact that one of the three blocked states relaxes much faster than the others, and thereby the charge measurement distinguishes two of the basis states from the rest of the states. By destructive measurement we mean that the entanglement is lost within a subspace, e.g., $a\ket{00}+b\ket{11} \rightarrow \ket{00}$ after observing an even outcome. We model this measurement protocol as a ZZ parity measurement followed by an initialization of the qubits.
Single-state discrimination refers to the case when the three blocked states remain in the original charge configuration and distinguished from the fourth state. For simplicity we take the discriminated state to be the $\ket{11}$ state an thereby the readout is effectively performed on the eigenbasis of the $CZ$ operator.

The main limitation of pairwise readout is that only one bit of information can be obtained from the two-qubit subspace. In the following we present three options to overcome this limitation and leverage the presence of the additional ancilla qubit.

The most straightforward way to overcome the limitations of the pairwise readout mechanisms of spin qubits is to utilize one of the two qubits as an ancilla in the usual sense and preserve the other one in the $\ket{0}$ state as a reference qubit for the readout as shown in Fig.~\ref{smfig_2ancilla_stabilizer} (top)~\cite{veldhorst2017silicon}. For most qubit layouts with low connectivity ($g\leq 4$) this necessitates additional SWAP gates between the ancillas, which in turn increases the susceptibility of the syndrome measurement to gate errors. However, we note that this protocol can be performed using CZ measurements and is also straightforwardly extendable to the case when a single particle is occupying the two dots and read out in the circuit QED protocol~\cite{burkard2021semiconductor}.

In order to reduce the number of additional SWAP gates in the measurement protocol, one could prepare the ancilla pair in the Bell-state
\begin{equation}
    \ket{\psi} = (\ket{00}+\ket{11})/\sqrt{2},
\end{equation}
and map the two halves of the plaquette operator to the two ancillas separately. As shown in Fig.~\ref{smfig_2ancilla_stabilizer} (middle) this protocol allows for stabilizer measurements without SWAP gates on the ancillas. One might expect that the same protocol would work with the preparation of the ancilla pair in the $\ket{00}$ state, but such a protocol is not fault tolerant as we will see later when discussing the measurements in the detecting region picture. The main limitation of the Bell-state approach is the fact that ancilla-initialization errors create two-qubit errors which typically significantly reduces the readout threshold. Furthermore, SWAP gates are still necessary in either X or Z type stabilizer measurements to avoid hook-error injection (discussed in Sec.~\ref{smsec_surface}).

Our third option of measuring stabilizers is relying on the concept of flag qubits \cite{Chamberland2018flagfaulttolerant}. Using an additional qubit as a flag qubit one can acquire additional information from the success of the stabilizer measurement process by monitoring Z errors in the part of the circuit where they can lead to hook errors. This flagging procedure needs to happen in the middle of the circuit by inserting CNOT gates between the flag and the ancilla qubits. Since we need to extract two bits of information from the ancilla pair, we need three subsequent (non-destructive) CZ measurements to distinguish the stabilizer and flag outcomes (see Fig.~\ref{smfig_2ancilla_stabilizer} (bottom)). Additionally, the flagged region cannot include SWAP gates because two-qubit error on the ancillas can induce a hook error and mute the flag at the same time. Due to these limitations and the increased circuit depth, we did not include the flag-qubit approach in our numerical analysis.

The syndrome measurement protocols can also be viewed in the detecting region picture (see App.~\ref{smsec:3CXdetregs} or Ref.~\cite{mcewen2023relaxing} for further information). If we start from the $\ket{00}$ state, two detecting region must appear, one corresponding to $Z_{a0}$ and one to $Z_{a1}$ (or any combination of these two). Let us consider a circuit without SWAP gates. Working with $Z_{a0}$ and $Z_{a0}Z_{a1}$ regions, the $Z_{a0}Z_{a1}$ region can be used to extended a four-body stabilizer as the blue region in the top right of Fig.~\ref{smfig_2ancilla_stabilizer}, but the region $Z_{a0}$ will expand only to the qubits on the upper end leading to a semi-plaquette stabilizer which does not commute with some other bulk stabilizers. On the other hand, working with a Bell pair implies the two ancilla stabilizers $X_{a0}X_{a1}$ and $Z_{a0}Z_{a1}$, where the X region will not be able to expand and therefore leads to the desired expansion of a plaquette stabilizer without any side effects. 

Similarly the SWAP syndrome measurement can be understood as a way of trying to avoid the expansion of the second stabilizer $Z_{a1}$, by swapping it with the other ancilla and not letting it participate in a two-qubit operation with the code qubits. The flag qubit approach extends the former concept by utilizing the second region to gather some error information from the syndrome ancilla during the circuit. This is achieved by expanding an X region to the syndrome ancilla which cannot expand to the code qubits but is sensitive to those $Z$ errors that would create a hook error by spreading to two of the code qubits.

\section{Surface code and the XZZX code}
\label{smsec_surface}

\begin{figure}
\includegraphics[width = .49\textwidth]{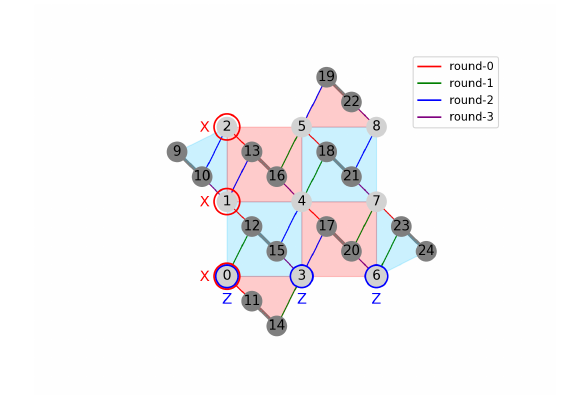}
\caption{Connectivity and schedule required for the simultaneous plaquette stabilizer measurements for a distance-3 surface code. Light (dark) gray nodes represent the code qubits (ancillas). Red (blue) circles denote qubits that participate in the $X$ ($Z$) logical operator with the corresponding single-qubit gate. $X$ ($Z$) plaquettes are shown in red (blue). For the final readout of the logical observable, an additional ancilla pair is required which can be placed next to qubit 0 or 8.}
\label{smfig_surface_schedule}
\end{figure}

Here we describe the properties of the surface code and the XZZX code and discuss some details of the numerical calculation. The surface code with the boundary conditions used in this work \cite{bombin2007optimal} is often referred to as the rotated rotated surface code which is a $[[d^2,1,d]]$ CSS stabilizer code \cite{calderbank1995good}. The XZZX code on the other hand is a variant of the surface code with locally rotated stabilizers featuring exceptionally high threshold against biased noise \cite{ataides2021xzzx}.

Improvements can be made in both cases, such as using a tailored decoder for biased noise for the surface code improving idling thresholds for high $\eta_T$ bias \cite{tuckett2020fault}, or choosing the boundary conditions of the XZZX code such that the high-rate errors have a longer path to logical failure \cite{ataides2021xzzx}. In this work, however, we did not explore such possibilities as they would have made the fair comparison of different codes difficult.

\subsection{Stabilizers and logical operators}

For both the surface code and the XZZX code the stabilizers are four- and two-body operators that can be divided into two groups as shown in red and blue on Fig.~\ref{smfig_surface_schedule}. For the regular surface code red (blue) plaquettes consist of only Pauli-X (Pauli-Z) while for the XZZX code each type of plaquette contains two X and two Z Paulis in the same orientation, e.g., $X_1Z_4Z_0X_3$ in the notation of Fig.~\ref{smfig_surface_schedule}. 

Logical operators are shown in Fig.~\ref{smfig_surface_schedule} for the $d=3$ surface code. Note that the X (Z) logical share a single qubit with only one two-body X (Z) plaquette on the boundary and two qubits with $(d-1)/2$ boundary plaquettes of the Z (X) type. Analogously, the logical X and Z operators in the XZZX code would be $\mathcal X^{(3)} = X_0Z_1X_2$ and $\mathcal Z^{(3)} = Z_0X_3Z_6$ in the $d=3$ case.

\subsection{Stabilizer measurements and CNOT scheduling}

The order in which we apply the CNOT gates is important for two reasons: {\it (i)} plaquette measurements have to commute and therefore if two plaquettes share two qubits, the schedule have to be such that the first plaquette gathers the parity information of the two qubits before a CNOT gate of the second plaquette would act on either of the qubits; {\it (ii)} the CNOT schedule of differently colored plaquettes must take the effect of hook errors into account~\cite{dennis2002topological}.

In the surface code a hook error is a pair of Pauli-X (Z) errors propagating from a faulty ancilla to the code qubits. In the case of $X$ ($Z$) plaquettes this can reduce the code distance of the logical $X$ ($Z$) operator depending on which qubits are affected \cite{dennis2002topological}. In order to avoid injection of hook errors, the schedule of the CNOT gates is chosen according to Fig.~\ref{smfig_surface_schedule}. Even though pairs of Pauli errors can still be injected by faulty ancillas, it still takes $(d+1)/2$ error events, to cause a logical error, as opposed to $\lceil (d+1)/4 \rceil$ for the schedule that induces hook errors. 

The schedule displayed in Fig.~\ref{smfig_surface_schedule} satisfies both the commutation and the hook error criteria. Note however that the layout does not favour the schedule for the red plaquettes. SWAP gates need to be applied on the respective ancilla pairs after round-0 and round-2 (which was not taken into account in Ref.~\cite{veldhorst2017silicon}). Since SWAP gates on the ancillas are inevitable due to the connectivity and the CNOT schedule, we used the SWAP-based protocol presented in Sec.~\ref{smsec_stab_meas}. The Z-plaquette measurement is shown in the top left of Fig.~\ref{smfig_2ancilla_stabilizer}. This only merely reduced the gate-error threshold, but allowed for a significant improvement for the readout threshold compared to the Bell-state approach.

\subsection{Fault-tolerant logical initialization and readout}

Here we briefly discuss the protocol for the fault-tolerant initialization and readout of an eigenstate of the logical Z operator of the surface code. The X observable of the surface code can be obtained by applying Hadamard gates on every code qubit in the beginning and before the final measurement, whereas the same protocol applies for the observables of the XZZX code with appropriate local basis transformations.

The key observation is that only X and Y errors can flip the eigenvalue of the logical Z operator, that are both detected by Z plaquettes. Therefore it suffices to know the eigenvalue of every Z plaquette in the first (last) round to ensure fault tolerance during the initialization (readout) of the logical observable. If every qubit is initialized in the $\ket{0}$ state in the beginning, the eigenvalue of the Z logical is $+1$ and every Z plaquette have eigenvalue $+1$. This means that one can already detect X and Y errors in the first round of stabilizer measurements by considering the Z plaquettes only. X-plaquettes on the other hand will have non-deterministic outcome in the first round, so they can only be compared with the previous round starting from the second measurement round.

The final measurement in the Z basis can be performed on a similar footing. After the last round of stabilizer measurements every qubit needs to be measured on the Z basis, from which one can infer the logical eigenvalue as well as every stabilizer eigenvalue to see if an error had happened right after the last round of stabilizer measurements that flipped the logical observable. For our spin-qubit lattice only ancillas can be read out directly, therefore, we need to swap every qubit with one of the nearby ancillas and measure the ancilla pairs afterwards. Note that in the case of the surface code, the final measurement requires only one extra pair of ancillas regardless of code distance.

The fault-tolerant initialization and readout of the logical observable is crucial not only for fault-tolerant computation on multiple logical qubits, but also in quantum memory experiments with a single logical qubit. If initialization and the final readout are assumed to be perfect, no logical error can appear on the logical observable in the presence of measurement errors only. This limit is clearly nonphysical and has significant impact on the overall error-threshold if the errors are biased towards readout errors as shown in Fig.~\ref{fig:optimal_readout}(a) for the case of the surface code.

\section{The 3-CX surface code}
\label{smsec:3CXdetregs}
\begin{figure}
\includegraphics[width = .49\textwidth]{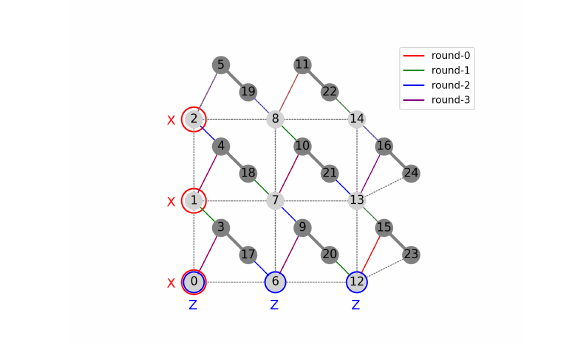}
\caption{Connectivity and schedule required for the simultaneous plaquette stabilizer measurements for the 3-CX surface code (distance-3). Light (dark) gray nodes represent the code qubits (ancillas). Red (blue) circles denote qubits that participate in the $X$ ($Z$) logical operator with the corresponding single-qubit gate. Red lines, representing CNOT gates in round-0 are covered by the purple lines in most cases.}
\label{smfig_3CX_surf_conn}
\end{figure}

The 3-CX surface code is a recently proposed variant of the surface code that relies on the concept of detecting regions in order to alleviate the connectivity requirements of the surface code~\cite{mcewen2023relaxing}. The spin-qubit connectivity graph is shown in Fig.~\ref{smfig_3CX_surf_conn}, where the hexagons and the squares of the surface code case are merged by removing one link per data qubit in the bulk. 

\subsection{Stabilizers and logical operators}

The stabilizer measurements are done in two half-cycles, where the CNOT schedule of the first half-cycle is shown on Fig.~\ref{smfig_3CX_surf_conn}. The second half-cycle uses the reversed schedule of the first one. Each of the half-cycles measure $d^2-1$ stabilizers. These are indeed the stabilizers of the surface code, but the $X$ and $Z$ type of bulk stabilizers are exchanged between the two half-cycles and the boundary stabilizers are adapting correspondingly (see Fig.~8.1 and 8.7 in Ref.~\cite{mcewen2023relaxing}). This choice also allows one to define the logical operators in the same way as for the rotated surface code.

Another important difference is the boundary of the physical qubit lattice, which is different from that of the rotated surface code, in order to preserve the code distance~\cite{mcewen2023relaxing}. A simple way to see this is trying to extend the structure in Fig.~\ref{smfig_3CX_surf_conn} downwards or to the left while preserving the connectivity pattern. This would mean adding ancilla pairs that are only connected to a single data qubit, preventing them from measuring multi-body stabilizers that are essential to protect the logical qubit.

\subsection{Stabilizer measurements in the detecting region picture}

A detecting region is a closed region of the circuit diagram originating from a reset (or set of resets) and terminating on a measurement (or set of measurements). A pure $X$ ($Z$) detecting region, one that only contains CNOT gates, can be defined by the property that a single Pauli $Y$ or $Z$ ($X$ or $Y$) error changes the total parity of the involved measurements. 

Detecting regions can be terminated in multiple ways on later measurements. This arbitrariness is the consequence of the fact that detecting regions at a given time are related to the generators of the spacetime stabilizer group at the corresponding time. As a simple example consider three qubits, where the parity of qubit $q_1$ and $q_2$ is measured via the ancilla $a$. The two data qubits are initially in the eigenstate of the $Z_1Z_2$ parity operator, which is the only stabilizer in the spacetime stabilizer group at $t=0$. Resetting the ancilla creates a single Z stabilizer $Z_a$ that commutes with the parity operator and therefore gets admitted in the stabilizer group at $t=1$~\cite{gottesman2022opportunities}. The choice of stabilizer generators is ambiguous, we can use e.g., $\{Z_a, Z_1Z_2\}$ or $\{Z_a, Z_aZ_1Z_2\}$.  We then apply a CNOT, controlled on $q_1$ and targeted on $a$, which extends the ancilla stabilizer as $Z_a \rightarrow Z_a Z_1$. For the two, mathematically equivalent, set of stabilizer generators we get $\{Z_a,Z_1Z_2\} \rightarrow \{Z_aZ_1,Z_1Z_2\}$ and $\{Z_a,Z_aZ_1Z_2\} \rightarrow \{Z_aZ_1,Z_a Z_2\}$.  One can regard a specific choice of generators as detecting regions. In the latter case the second detecting region is said to have contracted since its weight is reduced by one. Sticking to this example, after yet another CNOT between $q_2$ and $a$ we get $\{Z_aZ_1Z_2,Z_a\}$. Finally, measuring the ancilla removes $Z_a$ from the group as an element as well as from every element that contains it. Note that starting from a certain stabilizer $Z_1Z_2$ we contracted the corresponding detecting region and measured the parity. At the same time we expanded another region which took over the place of the original parity operator. Having a large number of qubits and stabilizers, the ambiguity of assigning resets to detecting regions creates several options for the tiling of detecting regions. One general design principle can be to maintain the $\mathbb Z_2$ anyon character of regions by ensuring that every qubit (in the bulk of the lattice) at every time is covered by a pair of $X$ and a pair of $Z$ detecting regions~\cite{mcewen2023relaxing}.

We have collected some effective rules for the construction of detecting regions in the case of the 3-CX surface code:
\begin{itemize}
\item[(i)] $Z$ ($X$) type regions can only originate/terminate on Z-basis ($X$-basis) resets/measurements;
\item[(ii)] if a $Z$ ($X$) type region is incident only on the target (control) qubit of a CNOT gate, it expands to the control (target) qubit;
\item[(iii)] if the same $Z$ ($X$) type region is incident on both the control and the target qubit of a CNOT gate the region contracts to the target (control) qubit.
\end{itemize}
For further information and rigour about detecting regions we refer the reader to Refs.~\cite{gottesman2022opportunities,mcewen2023relaxing}.

Between every measurement and reset the detecting regions represent the stabilizers of the surface code. The type of the boundary stabilizers does not change so the logical operators remain the same throughout the circuit. In the example of Fig.~\ref{smfig_3CX_surf_circ} the circuit of the $d=2$ version is shown at full width, using single (individually measurable) ancillas. We can see this as having a 4-body X plaquette and two 2-body Z plaquettes after the first measurement, but after the second measurement we see a 4-body Z plaquette expanding that will be the subject of our final measurement. The 2-body X plaquettes are not shown but using the rules above they can be straightforwardly identified as being located on in $q_0q_1$ and $q_2q_3$ edges, respectively.

\subsection{Fault-tolerant initialization and readout of the logical qubit}

\begin{figure*}
\includegraphics[width = \textwidth]{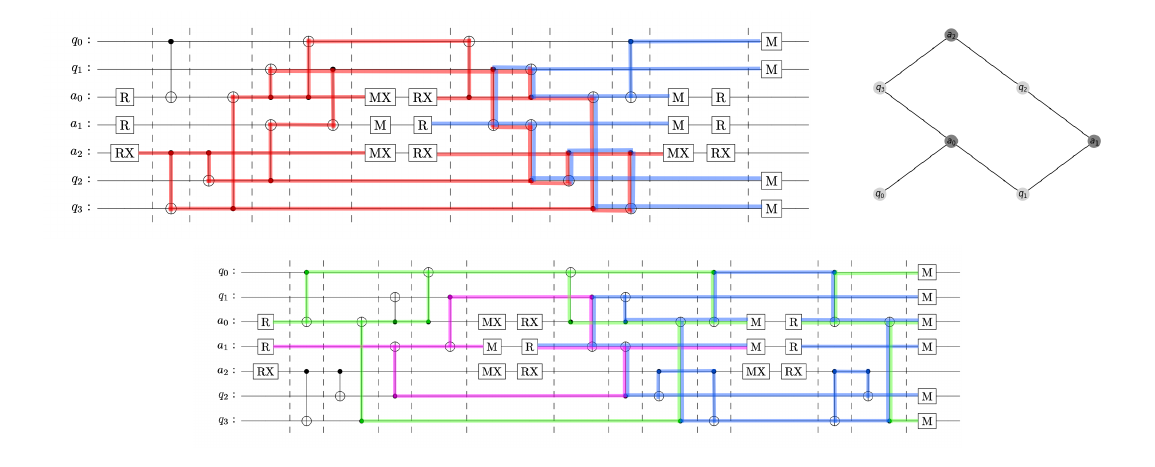}
\caption{Measurement circuit of the detecting regions and final measurement of the 3-CX surface code for $d=2$ with the required connectivity (top right). The two distinct detection rounds ending with ancilla measurements are alternating in the detection cycles. Top figure: an X type detecting region in a regular detection round (red), and the final measurement of the last Z type detecting region (blue) with $d^2$ measurements that is compatible with spin qubits. Bottom figure: two Z type regions in a regular detection round (green and purple), and final detection round with three regions (blue and green) terminated on $\sim 2d^2$ measurements as in Ref.~\cite{mcewen2023github}.  Dashed lines separate different layers of CNOT gates, simplifying the comparison with the time-slice picture of Ref.~\cite{mcewen2023relaxing}.}
\label{smfig_3CX_surf_circ}
\end{figure*}

For the initialization we can follow directly Ref.~\cite{mcewen2023github}, but we need to modify the final measurement part, since the solution provided by McEwen {\it et al.} measures every qubit (both data and ancilla) in the half-cycle state. This is not possible in spin qubit architectures, since qubits can only be measured pairwise, and only ancilla types have reference qubits for the measurements. In the following we present a scheme for the final measurement round in which only the data qubits need to be measured. For spin qubits this can be done in the same way as for the rotated surface code, by swapping one of the ancillas from each pair with a neighbouring data qubit and measuring the ancilla-data pair.

As described in Ref.~\cite{mcewen2023relaxing} each detecting region has to be terminated on an appropriate measurement. However, as for the conventional rotated surface code, only those stabilizers play a role in the decoding which are sensitive to errors that affect the outcome of the logical measurement. Consequently, one can measure every data qubit in the Z basis right after the last round of syndrome measurements as in the upper part of Fig.~\ref{smfig_3CX_surf_circ}. From this the Z eigenvalue of the logical qubit can be inferred and every Z-type detecting region is terminated. The syndromes obtained for each detecting region can be compared to the appropriate values of the last syndrome detection round the same way as if the final measurement would have been round $(T+1)$. 

For comparison the final measurement scheme of Ref.~\cite{mcewen2023github} is shown in the lower part of Fig.~\ref{smfig_3CX_surf_circ}. After the last two CNOT layers, the stabilizer group is extended with a trivial element $Z_{a1}$ and a nontrivial element $Z_{a0}Z_{q_0}Z_{q3}$ necessitating the measurement of ancillas $a_0$ ans $a_1$. In this scheme in total $2d^2-1-\lceil (d-1)/2 \rceil$ measurements are required (of which only $\mathcal O(d)$ are trivial), but in the protocol we used in this work only $d^2$ simultaneous measurements are needed in the final round.

\section{The XYZ$^2$ matching code}
\label{smsec:xyz2}
\begin{figure}
\includegraphics[width = .49\textwidth]{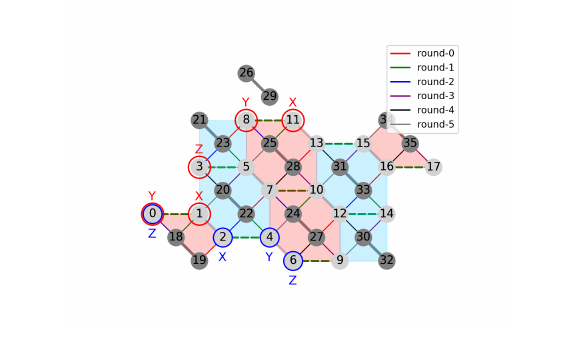}
\caption{Connectivity and schedule required for the simultaneous plaquette stabilizer measurements for a distance-3 XYZ$^2$ matching code. Light (dark) gray nodes represent the code qubits (ancillas). Green dashed line shows the qubit pairs along which the link stabilizers are measured. Red (blue) circles denote qubits that participate in the $Y$ ($Z$) logical operator with the corresponding single-qubit gate.}
\label{smfig_XYZ2_schedule}
\end{figure}

The XYZ$^2$ code is a $[[2d^2,1,d]]$ stabilizer code~\cite{srivastava2022xyz} defined on a hexagonal lattice. This code is a special case of the hexagonal matching code family~\cite{Wootton_2015}, with 2-body link operators and appropriate boundary conditions. The connectivity required for this code can be embedded in a square grid of spin qubits, where ancilla pairs are located in the middle of each hexagon, giving the highest encoding density $\sim 1$ qubit/ancilla among the considered error correction codes.

\subsection{Stabilizers and logical operators}

In the hexagonal matching code there are two types of stabilizers, plaquettes and links. There are $d^2$ link operators which are products of Pauli-Z operators for two neighboring qubits, where the neighbors are to be taken in the same direction for every pair (see dashed green lines on Fig.~\ref{smfig_XYZ2_schedule}). Plaquettes in the bulk are six-body Pauli operators of alternating type, i.e., starting with a Pauli-Z on a qubit of the hexagon where a link is incident, followed by X, Y, Z, X and Y operators in a clockwise direction along the hexagon. In total there are $(d-1)^2$ bulk plaquettes (blue and red hexagons on Fig.~\ref{smfig_XYZ2_schedule}). At the boundary of the lattice one finds $2d-2$ semiplaquettes which are three-body Pauli operators defined by the same rule as their bulk counterparts~\cite{srivastava2022xyz}.

As in the case of the surface code, plaquette operators can be divided into two groups as shown in Fig.~\ref{smfig_XYZ2_schedule}. Such a division will be useful for the CNOT scheduling of plaquette measurements as well as for the matching decoder developed for this code.

Logical operators are defined on the edges of the diamond-shaped boundary and contain both X, Y, and Z Paulis as shown on Fig.~\ref{smfig_XYZ2_schedule}. This choice differs from that of Ref.~\cite{srivastava2022xyz} only by stabilizer operators. The motivation behind defining them on the boundary comes from the similarity to the surface code which allows one to perform the decoding in an analogous way.

\subsection{Stabilizer measurements and CNOT scheduling}

Measurements of plaquette stabilizers can be performed via a pair of ancillas sitting in the middle of each plaquette as in Fig.~\ref{smfig_XYZ2_schedule}, resulting in a connectivity requirement of a simple square lattice. Link measurements reuse the ancilla pairs of the plaquettes and therefore performed in separate measurement rounds, after the plaquette measurements.

\begin{figure}
\includegraphics[width = 0.49\textwidth]{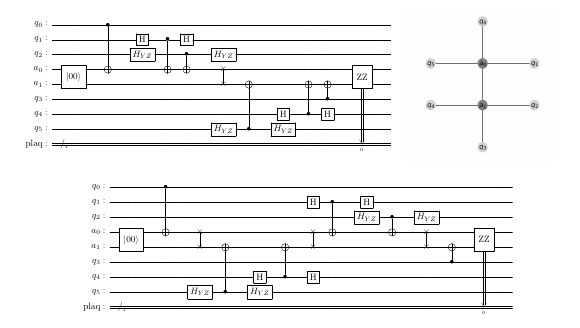}
\caption{Top (bottom) figure presents the stabilizer circuit of the blue (red) plaquettes of Fig.~\ref{smfig_XYZ2_schedule} requiring one (three) SWAP gates on the ancilla pairs. The labeling of the data and ancilla qubits starts with the qubit that is acted on in round-0 for both type of plaquettes, and continues clockwise.}
\label{smfig_XYZ2_plaquette_circuit}
\end{figure}

The specific choice for the layout in Fig.~\ref{smfig_XYZ2_schedule} ensures, that link stabilizers can be measured without using any SWAP gates. On the other hand, plaquette measurements do require SWAP gates (one for blue plaquettes three for reds) as shown on Fig.~\ref{smfig_XYZ2_plaquette_circuit} and a CNOT schedule that allows for neighboring plaquette measurements to commute and avoids injecting hook errors. In this work the schedule was chosen according to Fig.~\ref{smfig_XYZ2_schedule}. Further details on the error chains created by single error events and the corresponding syndromes will be discussed in the next section.

\subsection{Decoding}

\begin{figure}
\centering
\includegraphics[width = .49\textwidth]{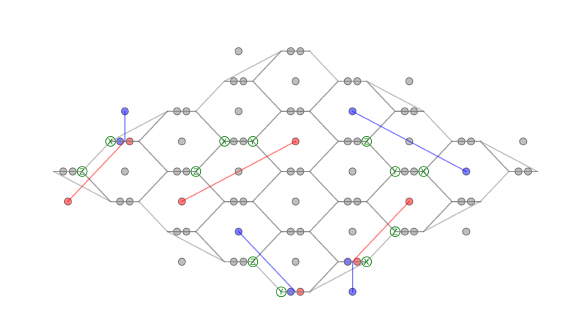}
\caption{Schematic figure of a $d=5$ matching code with the scheduling chosen according to Fig.~\ref{smfig_XYZ2_schedule}. Y and Z errors on the ancilla qubit lead to one-, two-, and three-qubit errors. These error-chains correspond to pairs of detection events in the two sublattices. Two- and three-qubit errors propagate parallel to the respective edge, therefore cannot induce logical errors. Green X,Y, and Z denote the type of error occurring on the data qubits. Plaquette detection events on the two different type of plaquettes are shown as blue and red nodes, while link detection events are shown as a pair of blue and red nodes, referring to their role in both decoding problems. Gray nodes show stabilizers where no error was detected.}
\label{smfig_xyz2_decoding}
\end{figure}

Even with the careful scheduling of the CNOT gates, single errors on the ancillas can lead to syndromes with up to five detection events. E.g., a Y error on the ancilla (not the reference qubit) flips the measurement outcome, which flips back again in the next round (two detection events). If such an error happens after round-1 or round-3 it induces Pauli errors on two data qubits which change the eigenvalue of one link and two plaquette stabilizers (see examples in Fig.~\ref{smfig_xyz2_decoding}). 

However, same-coloured plaquettes, combined with every link stabilizers, can be gathered into subgraphs and decoded separately. For a given subgraph, the single-error syndromes become analogous to the case of the surface code, with each syndrome containing at most four detection events. This is the direct consequence of the matching code family being a $D(\mathbb Z_2)$ anyon model~\cite{Wootton_2015}, where the plaquette operators host anyons of the corresponding color and links host fermions, i.e., a pair of red and blue anyons. When we separate the two subgraphs, the link fermions are divided into two anyons.

Four-body detection events can be decomposed into a time-like pair, i.e., measurement error, and a space-like pair, i.e., data qubit error or error-chain. Note that space-like pairs do not necessarily appear at the same time, due to the scheduling of plaquette measurements and the separately measured links. Nevertheless, time-like pairs can be separated easily since they are always local in space.

In the decoding graph nodes correspond to possible detection events (i.e., comparison of subsequent stabilizer measurements) and edges connect two events that can appear simultaneously as a result of a single error. We can take advantage of the decomposition of time-like and space-like syndrome pairs by assigning a weight to every edge based on the joint probability $p$ of all the single-qubit errors that result in the corresponding pair of syndromes. The weights are calculated from the probabilities as $w = \log(p/(1-p))$. Given a weighted decoding graph, a perfect matching decoder can be used to find an error configuration that is consistent with the syndromes.

In this work we assigned the error probabilities in the following way:
\begin{widetext}
\begin{eqnarray}
\begin{split}
p^{1q}_{Z} = \sum \limits_{i+j+k+l+m \in \text{odd}} {2 \choose i} \pTt^i(1-\pTt)^{2-i} {4 \choose j} (\pGo/3)^j(1-\pGo/3)^{4-j} {4 \choose k} (4\pGt/15)^k(1-4\pGt/15)^{4-k} \times \\
 {N_2 \choose l} (8\pGt/15)^l(1-8\pGt/15)^{N_2-l} (4\pGt/15)^m(1-4\pGt/15)^{1-m} \, ,
\end{split}
\end{eqnarray}
\end{widetext}
is the probability of a Z-error on a single qubit, since every qubit is subject to two idling phase during measurements, four single qubit gates (from two plaquettes), and four two-qubit gates (three from plaquettes, one from a link), and finally a two-qubit gate error inducing a Y or Z ancilla error that propagates to the qubit as a Z error (can happen via one of the plaquettes for every qubit or via the link for every second qubit). In the second row $N_2$ denotes the total number of two-qubit gates in round-0 and round-4 of the plaquette measurement, i.e., $N_2=2$ for red plaquettes $N_2=4$ for blues (see also in Fig.~\ref{smfig_XYZ2_plaquette_circuit}). A Pauli-Z error on a data qubit creates a pair of detection events on two neighboring plaquettes of the same type.

Similarly the probability of an X or Y error is
\begin{widetext}
\begin{eqnarray}
\begin{split}
p^{1q}_{XY} = \sum \limits_{i+j+k \in \text{odd}} {2 \choose i} \pTo^i(1-\pTo)^{2-i} {4 \choose j} (2\pGo/3)^j(1-2\pGo/3)^{4-j} {4 \choose k} (8\pGt/15)^k(1-8\pGt/15)^{4-k} \, .
\end{split}
\end{eqnarray}
\end{widetext}
Such an error induces detection events on a link and on two plaquettes of different types.

Furthermore, the probability of a time-like error, i.e., measurement error is
\begin{widetext}
\begin{eqnarray}
\begin{split}
p^{m} = \sum \limits_{i+j+k \in \text{odd}} \pR^i(1-\pR)^{1-i} {2 \choose j} (8\pR/15)^j(1-8\pR/15)^{2-j} {N_2 \choose k} (8\pGt/15)^k(1-8\pGt/15)^{N_2-k} \, .
\end{split}
\end{eqnarray}
\end{widetext}
where the first term is the readout assignment error and the second corresponds to an error in the measurement or the initialization of the qubit. This time $N_2$ is the number of all two-qubit gates ($N_2 = 7$ for red plaquettes and $N_2=9$ for blue plaquettes and $N_2 = 2$ for links). The set of errors that occur with this probability create two detection events at the same plaquette or link in two subsequent measurement cycles.

Finally, the probability of a faulty ancilla injecting two or three Pauli errors is more straightforward to calculate as there are only two possibilities for an ancilla to pick up a Y or Z error (after round-1 or round-3 for the two-qubit case and during or after round-2 for the three-qubit case). Therefore we get 
\begin{eqnarray}
p^{2q} = p^{3q} = 2(8\pGt/15)(1-8\pGt/15)\, ,
\end{eqnarray}
for the two- and three-qubit error-chain probabilities. The syndromes induced by such error chains are less straightforward, therefore we summarize them in Fig.~\ref{smfig_xyz2_decoding}. Importantly none of these reduce the code distance and both of them create a pair of (relevant) detection events.

Close to the boundaries, single detection events can appear that must be connected to a boundary node. There are in total four boundary nodes, two for each type of plaquettes. If the single detection event is associated to that boundary where one of the logical operator is defined, the graph edge connecting the single node to the boundary node will also imply a logical error.

\subsection{Fault-tolerant initialization and readout of the logical qubit}

\begin{figure}
\includegraphics[width = .49\textwidth]{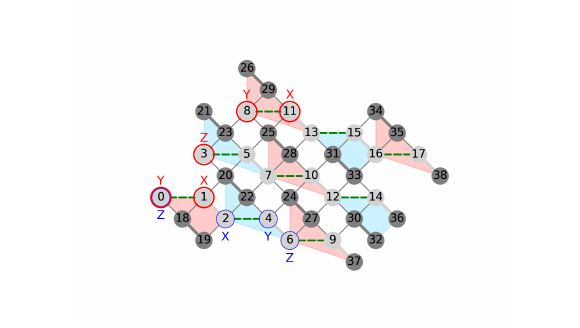}
\caption{Full connectivity map, extended with $d$ ancillas (labelled 37-38) for the logical initialization and readout. Qubits are divided into overlapping groups of five, of which two pairs of qubits can be read out simultaneously (denoted by thick light- and dark gray links).}
\label{smfig_XYZ2_noisyLogRO}
\end{figure}

The hexagonal matching code is the only candidate in our work that utilizes one qubit per ancilla for the encoding. Due to the additional constraint on the readout (pairwise, destructive), it is a nontrivial task to initialize and readout the logical qubit while inferring sufficiently many stabilizer eigenvalues to maintain fault tolerance. Here we describe in detail the scheme for the logical initialization and readout used in this work.

Since every qubit is participating in one link only, one shall consider all the links and distribute them into two groups such that red (blue) links are the ones matching two non-neighboring blue (red) plaquettes on their corners. Regarding the red (blue) logical, i.e., logical Y (Z), the relevant stabilizer eigenvalues for decoding are the plaquettes colored red (blue) on Fig.~\ref{smfig_XYZ2_schedule} as well as every link. To initialize or measure all these stabilizer eigenvalues and the logical qubit on the Y (Z) basis, we initialize or measure red (blue) links on the $XY$-$YX$ basis and blue (red) links on the $ZI$-$IZ$ basis. Knowing both $m_{XY}$ and $m_{YX}$ eigenvalues at a given step (initial or final), the eigenvalue of the corresponding Z-link is just the product of the two, i.e., $m_{ZZ} = m_{XY} m_{YX}$, (similarly for the $ZI-IZ$ case) providing every link stabilizer eigenvalue. Furthermore, the relevant plaquette eigenvalues can be constructed from products of four known eigenvalues ($ m_{XY} $ and $ m_{YX} $ on the neighboring links as well as $ m_{ZI} $ and $ m_{IZ} $ on the incident links). Finally, one can easily check, that the logical eigenvalue is also determined by the appropriate product of the aforementioned link eigenvalues (see Fig.~\ref{smfig_XYZ2_noisyLogRO}).

\begin{figure}
\includegraphics[width = 0.49\textwidth]{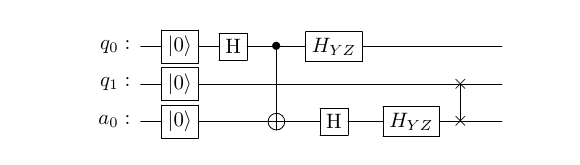}
\caption{Initialization of the $XY$-$YX$ link operator in the mutual $+1$ eigenstate. Initialization of the $ZI$-$IZ$ link is trivial after initialization of every qubit pair in the $\ket{00}$ state.}
\label{smfig_XYZ2_Loginit_circuit}
\end{figure}

\begin{figure}
\includegraphics[width = .49\textwidth]{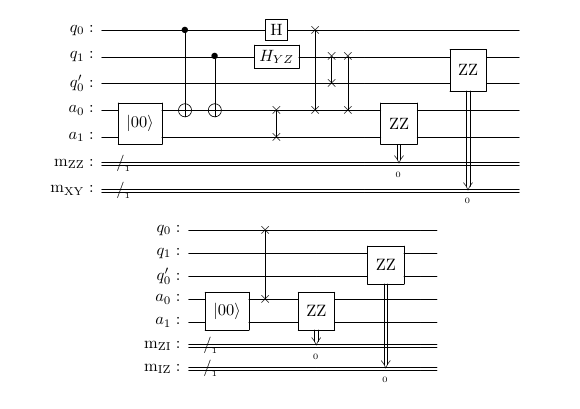}
\caption{Top: final readout circuit for the $XY$-$ZZ$ (or equivalently $XY$-$YX$) link measurement. Bottom: $ZI$-$IZ$ link measurement. The SWAP gate between $q_0$ and $a_0$ in both circuits ensures that $q_0'$ (i.e., $q_0$ of the neighboring group) is in the $\ket{0}$ state before the pairwise readout.}
\label{smfig_XYZ2_LogRO_circuit}
\end{figure}

In order to perform the required $XY-YX$ and $ZI-IZ$ link measurements, we append $d$ additional ancillas and regroup the data and ancilla qubits as shown in Fig.~\ref{smfig_XYZ2_noisyLogRO}. Despite the different shapes (triangle or tetragon) and types (red or blue), every group has at least the following connectivity: two qubits, $q_0$ and $q_1$, are connected to the same ancilla $a_0$ which can be read out or initialized together with another ancilla $a_1$; a third qubit $q_0'$ (that is in state $\ket{0}$ or acts as $q_0$ in a neighboring group) can be read out or initialized together with $q_1$. Figs.~\ref{smfig_XYZ2_Loginit_circuit}-\ref{smfig_XYZ2_LogRO_circuit} show the circuits for the initialization and readout of such link operators.

\section{Floquet codes}
\label{smsec:floquet}
\begin{figure}
\includegraphics[width = .49\textwidth]{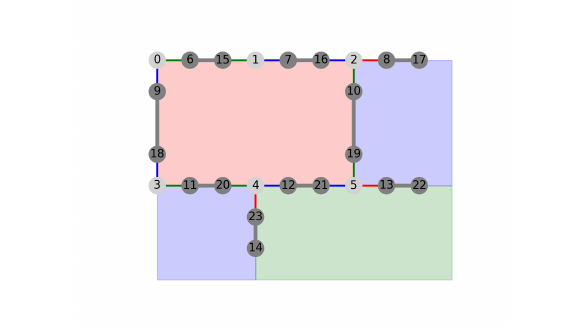}
\caption{Connectivity required by the Floquet codes ($d=2$), with tricoloring of the hexagons and the links according to the color code. Convention X: red, Y: green, Z: blue.}
\label{smfig_floquet_conn}
\end{figure}

Floquet codes on a honeycomb lattice can be derived from the color code in the anyon condensation picture~\cite{kesselring2022anyon}. The color code is defined on the honeycomb lattice, where the vertices correspond to the data qubits, and the stabilizers are six-body $X$ and $Z$ plaquette operators on the hexagons that are divided into three color groups such that each hexagon is surrounded by contrasting colors. In this model nine bosonic excitation can be defined labeled as red, green, blue and $x$, $y$, $z$. These bosons have non-trivial braiding statistics with each other if they have different color and Pauli labels, thus we refer to them as {\it anyons}.

In the Floquet codes under consideration, the nine bosons of the color code are condensed into two set of bosons that braid non-trivially with one another~\cite{kesselring2022anyon}. The resulting code is therefore part of the surface code family, i.e., a $D(\mathbb Z_2)$ anyon model. By periodically changing which anyon is condensed, an arbitrary single-qubit error can be detected on the honeycomb lattice while the logical qubit remains well-defined. The change in the anyon condensation should always be done in a way that one of the $e$ ($m$) anyons from the old code remains an $e$ ($m$) anyon in the new condensed code. 

The main difference between the two Floquet codes considered in this work, the Floquet color code~\cite{kesselring2022anyon} and the honeycomb Floquet code~\cite{Hastings2021dynamically}, is which of the condensed phases are used and how they are transformed to one another in subsequent measurement rounds. A noteworthy difference is that the Floquet color code has been constructed such that it is a CSS code.

\subsection{Stabilizers and logical operators}

As discussed above, on the abstract level a condensed color code at a given time is equivalent to a surface code. However on the microscopic level the relation between detection cells (detection cells in the terminology of Ref.~\cite{kesselring2022anyon} are detecting regions where the stabilizer measurements are considered as a single step) and anyons is different. This is due to the fact that the condensed color code has two different Pauli stabilizers on a given plaquette, as opposed to the $D(\mathbb Z_2)$ anyon models we studied before. Stabilizer measurements in the Floquet code are performed by measuring a set of $XX$, $YY$, or $ZZ$ links (in the Floquet color code only $XX$ and $ZZ$) that are connecting same coloured plaquettes in subsequent measurement rounds. The links can be colored as well according to the plaquettes they connect.

Let us consider the point in time when $X$ links are measured on the red edges (see Fig.~\ref{smfig_floquet_conn}). Each link will anticommute with the red $Z$ plaquettes, which we therefore must have measured in the previous round to ensure fault tolerance (every region lives for four rounds in the 6-round-long measurement cycle). The link measurements allow us to infer the green and blue $X$ plaquettes, which marks the end of the green $X$ detection cells and the beginning of a blue $X$ cells. In total we have four detecting regions after the red $X$ link measurements: blue $X$, blue $Z$, red $X$, and green $Z$. If a Z error happens somewhere a red and a green $X$ region will report it. Similarly an X error shows up on green and blue $Z$ regions. Therefore we can identify red and a blue $X$ plaquettes with non-trivial outcome in this round with $e$ anyons, while red and green $Z$ plaquettes will host $m$ anyons. 

Logical operators also transform together with the stabilizer group. The transformation of the logicals can be viewed as follows: having a logical operator defined in a given phase, e.g., a string of Pauli operators, one can define an equivalent logical operator by multiplying it with all the link stabilizers that have support on the logical, thereby shifting the logical or changing its Pauli type. One can choose the next phase such that it is compatible with the one of the equivalent versions of the logical operators ensuring the conservation of the logical subspace~\cite{kesselring2022anyon}.

\subsection{Details for the spin qubit version}

For both of the Floquet codes we considered periodic spatial- and open temporal boundary conditions. Since the plaquette stabilizers are not measured in a single round, the effectively perfect logical initialization/measurement does not lead to such a side-effect as for the static stabilizer codes. For both versions given a code distance $d$ we performed $d/2$ measurement cycles, that is $3d$ measurement rounds.

Furthermore, we used the Bell-state protocol of Sec.~\ref{smsec_stab_meas} for the link measurements. Since the link measurements use two CNOT gates, removing the SWAP gate from the ancillas considerably increases the gate error threshold (from about $0.35\%$ to $\sim 0.45\%$). On the other hand, the multi-round measurement cycle results in a low readout threshold ($\pR^\text{th} \sim 1\%$) for any of the stabilizer measurement protocols, therefore the link-error injection due to faulty ancilla initialization does not lead to such a significant compromise as for the static stabilizer codes.

Finally we note that only one logical observable was used for the threshold calculations, but we tested two different axis for the idling bias to avoid getting an overly optimistic threshold for a given logical, as typical for CSS codes.

\subsection{Comparison of the Floquet color code and the honeycomb Floquet code}

\begin{figure*}
    \includegraphics[width=\textwidth]{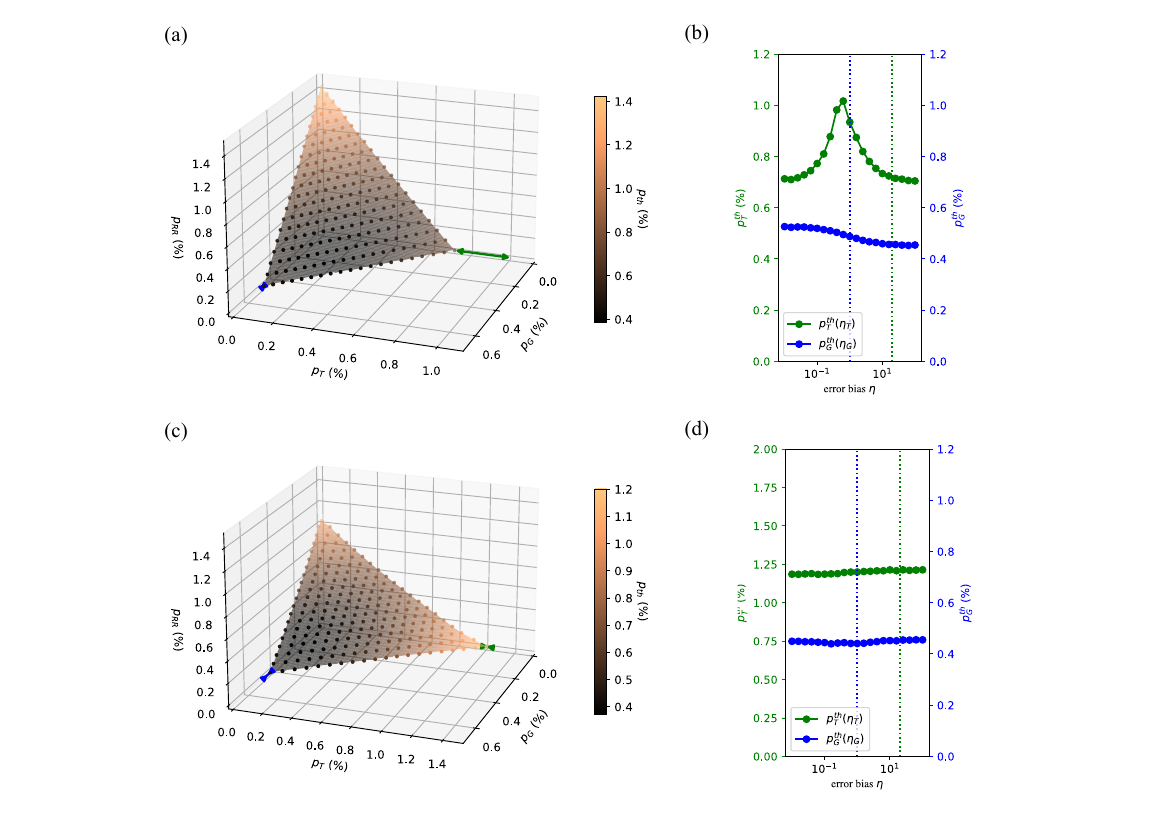}
    \caption{(a) Threshold surface of the Floquet color code for $\eta_T = 20$ and $\eta_G =1$, (b) Dependence of $\pth_G$ on the gate error bias $\eta_G$ and $\pth_T$ on the idling error bias $\eta_T$ for Floquet color code. Each point is calculated from code distances $d\in \{10,12,14,16\}$ and $T=d/2$ stabilizer cycles (3 rounds per cycle). (c)-(d) Similarly, threshold surface and bias dependencies of the honeycomb Floquet code.}
\label{smfig_honeycomb_vs color_floquet}
\end{figure*}

A side-by-side comparison of the threshold parameters for the Floquet color code and the honeycomb Floquet code is shown in Fig.~\ref{smfig_honeycomb_vs color_floquet}. Both threshold surfaces are very well approximated by a plane and have similar threshold parameters. The honeycomb code, as opposed to the Floquet color code, has a remarkably small sensitivity to noise biases both for gate errors and idling errors.

\section{Details of the numerics}

For the Clifford-circuit simulations we used stim~\cite{Gidney2021stim}. For each code distance and physical error rate we took up to $300\,000$ shots unless $30\,000$ logical failures are encountered before that. Decoding graphs are built based on the respective stim detector error models, except for the XYZ$^2$ code which is detailed in App.~\ref{smsec:xyz2}. The reason for the latter was that stim was unable to decompose some of the hyperedges in the decoding graph. The decoding was performed with pymatching in each case~\cite{higgott2022pymatching}.

\bibliography{references}

\end{document}